\documentclass[12pt,preprint]{aastex}
\usepackage{graphicx}
\usepackage{amssymb}
\usepackage{amsmath}
\usepackage{subfigure}
\usepackage{multirow}
\usepackage{threeparttable}
\usepackage{array}
\usepackage{natbib}
\usepackage{lscape}

\newcommand{\reffig}[1]{Fig.~\ref{#1}}
\newcommand{\reftable}[1]{Table \ref{#1}}

\begin{document}
\title{The short bursts in \object{SGR 1806-20}, \object{1E 1048-5937}  and \object{SGR 0501+4516}}

%\author{Zhijie Qu\altaffilmark{1}, Zhaosheng Li\altaffilmark{1}, Yupeng Chen\altaffilmark{2}, Shi Dai\altaffilmark{1}, Long Ji\altaffilmark{2}, Renxin Xu\altaffilmark{1,3}, Shu Zhang\altaffilmark{2}}

\author{Zhijie Qu\altaffilmark{1}}
\email{quzhijie@pku.edu}

\author{Zhaosheng Li\altaffilmark{1}}
\email{lizhaosheng@pku.edu.cn}

\author{Yupeng Chen\altaffilmark{2}}
\email{chenyp@ihep.ac.cn}

\author{Shi Dai\altaffilmark{1}}
\email{dash41198@gmail.com}

\author{Long Ji\altaffilmark{2}}
\email{jilong@ihep.ac.cn}

\author{Renxin Xu\altaffilmark{1,3}}
\email{r.x.xu@pku.edu.cn}

\author{Shu Zhang\altaffilmark{2}}
\email{szhang@ihep.ac.cn}

\altaffiltext{1}{School of Physics and State Key Laboratory of Nuclear Physics and Technology, Peking University, Beijing 100871, P. R. China}
\altaffiltext{2}{Institute of High Energy Physics \& Theoretical Physics Center for Science Facilities, Chinese Academy of Sciences, Beijing 100049, China}
\altaffiltext{3}{Kavli Institute for Astronomy and Astrophysics, Peking University, Beijing 100871, P. R. China}

\begin{abstract}
We analyzed temporal and spectral properties, focusing on the short bursts, for three anomalous X-ray pulsars (AXPs) and soft gamma repeaters (SGRs), including \object{SGR 1806-20}, \object{1E 1048-5937} and \object{SGR 0501+4516}. Using the data from \textit{XMM-Newton}, we located the short bursts by Bayesian blocks algorithm. The short bursts' duration distributions for three sources were fitted by two lognormal functions. The spectra of shorter bursts ($< 0.2~\rm s$) and longer bursts ($\geq 0.2~\rm s$) can be well fitted in two blackbody components model or optically thin thermal bremsstrahlung model for \object{SGR 0501+4516}. We also found that there is a positive correlation between the burst luminosity and the persistent luminosity with a power law index $\gamma = 1.23 \pm 0.18 $. The energy ratio of this persistent emission to the time averaged short bursts is in the range of $10 - 10^3$, being comparable to the case in Type \uppercase\expandafter{\romannumeral1} X-ray burst.
\end{abstract}

\keywords{Stars}

\maketitle

\section{Introduction}

Anomalous X-ray pulsars (AXPs) and soft gamma repeaters (SGRs) are isolated neutron stars, now regarded as ``magnetars". As X-ray pulsars, their rotation periods vary from $\sim 2$ to $\sim 10\rm ~s$, while spin-down rates cover $ 10^{-13} - 10^{-11} \rm ~ s~s^{-1}$ \citep{Mereghetti2008}. Except for some strange magnetars (e.g. \object{SGR 0418+5729}, \citealt{Rea2013}), both of these two parameters are larger than in normal radio pulsars, which results in an ultra-strong magnetic field exceeding the quantum critical value ($B_{\rm{QED}} = 4.4 \times 10^{13} ~ \rm{G}$) in AXPs/SGRs. Here, the assumption is used that the AXPs/SGRs are braked by magnetic dipoles in a vacuum. During outburst, their persistent soft X-ray luminosity ($\sim 10^{34}-10^{36}~\rm erg~s^{-1}$) usually exceeds their rotational energy loss rates ($\sim 10^{33} \rm~erg~s^{-1}$) \citep{Mereghetti2008}.  This characteristic is considered as an important boundary between magnetar and normal pulsars. However, the discovery of \object{PSR J1846-0258} blurred this boundary as this object has magnetar-like bursts and a persistent X-ray luminosity comparable with its rotational energy loss rate \citep{Gavriil2008}. AXPs/SGRs also have temporal activities with different time scales, such as glitches/anti-glitches (lasting several dozen days, including the recovery stage, \citealt{Archibald2013}), outburst (lasting several months to years) and short burst (lasting $\sim 0.1~\rm s$).

\citet{Duncan1992} first presented the ``magnetar" concept and discussed the formation of a magnetar. They suggested that an $\alpha\Omega$ dynamo operating in a neutron star with initial period $P \sim 1 \rm~ ms $ could generate a dipole magnetic field much stronger than $10^{13}~\rm G$. \citet{Thompson1995} regarded SGRs as a class of magnetar and suggested that the large scale reconnection or instability of magnetic field could account for short bursts and the giant flare in \object{SGR 0526-66}. \citet{Thompson1996} considered the diffusive crust activity producing low amplitude Alfv{\' e}n waves in the magnetosphere as an effective way to transfer the magnetic energy into persistent X-ray emission. \citet{Kouveliotou1998} measured the spin-down rate of \object{SGR 1806-20} and confirmed the ultra-strong magnetic field in the dipole magnetic field assumption. \citet{Kouveliotou1998}, combined with results from subsequent papers (e.g. \citealt{Marsden1999}, \citealt{Dib2009}) on other sources, was regarded as substantial evidences for magnetar model. \citet{Lyubarsky2005} considered magnetic reconnection in relativistic treatment within magnetar framework. \citet{Perna2011} did a quantitative simulation to trigger the short bursts based on a starquake caused by the breaking of neutron star crust. However, some challenges inevitably appeared, such as the existence of low magnetic field magnetars (\object{SGR 0418+5729}, \citealt{Rea2013}; \object{Swift J1822.3−1606}, \citealt{Rea2012}; \object{3XMM J185246.6+003317}, \citealt{Zhou2014}, \citealt{Rea2014}), and the predictions that magnetars should have large spatial velocities and energetic-associated supernovae \citep{Duncan1992}, neither of which has been observed yet \citep{Vink2006, Mereghetti2008}.

\citet{Chatterjee2000} developed an accretion disk model for AXPs/SGRs, whereby the emission is powered by accretion from a fossil disk. \citet{Wang2006} deduced there might be a disk around \object{4U 0142+61} from the spectral-energy distribution in the optical/infrared band, where the disk may come from a supernova fallback. The accretion-based models were usually criticized because of the lack of a mechanism to explain the giant flares and bursts. Thus, these models require input from the magnetar model to become a ``hybrid" and complete model \citep{Mereghetti2008}. Nevertheless, combining the accretion model and strange matter state, Xu and coworkers \citep{Xu2003, Zhou2004, Xu2006, Xu2007} suggested that the solid quark stars, instead of neutron stars, could generate giant flares and bursts in the process of accretion-induced star-quakes. Massive white dwarfs with larger rotational energy release than neutron stars are also regarded as an alternative model for AXPs/SGRs \citep{Malheiro2012}.

Comparing the magnetar model and the accretion model, the main difference is the origin of energy. Magnetic energy release is responsible for the persistent and burst radiation in the magnetar model, while it is gravitational energy of accreted matter or elastic energy of solid matter \citep{Zhou2014a} which produces this emission in the accretion model. One way to distinguish the mechanisms of persistent and burst radiation is analyzing the spectra, including the continuum and emission or absorption lines. \citet{Ibrahim2003} and \citet{Ibrahim2007} did a series of studies of spectral features at $\sim 5~\rm keV$ and $\sim 20~\rm keV$ from \object{SGR 1806-20}. They regarded these features as evidence for the proton-cyclotron resonance (PCR), which indicates that the surface magnetic field could reach $\sim 10^{15}\rm ~ G$. \citet{Bernardini2009} found a spectral feature at $\sim 1.1\rm~ keV$ in the AXP \object{XTE J1810-197} which requires a $\sim 10^{14}\rm~G$ magnetic field if it is from PCR. \citet{Tiengo2013} discovered a phase-dependent feature with a ``V'' shape in the phase-resolved, persistent spectrum of \object{SGR 0418+5729}, and they interpreted this result as evidence for a twisted magnetic field. \citet{Vigano2014} showed that the spectral features in the thermally dominated range could also be the result of inhomogeneous surface temperatures, without any dependence on the magnetic field. However, this interpretation does not adequately describe the phase-dependent feature of \object{SGR 0418+5729}.

The uncertainty in determining the emission mechanisms from line features arises mainly because the observations are not sufficient to distinguish between the theoretical models. Thus, previous studies have focused on the the interpretation of continuum spectra since these data are better able to constrain the models\citep{Fenimore1994}. With this fact in mind, \citet{Nakagawa2011} and \citet{Enoto2012} studied the continuum of persistent radiation and weak burst spectra of \object{SGR J0501+4516} and \object{SGR J1550-5418} from \textit{Suzaku} observations. They found these spectra to have similar shapes, and thus they claimed the persistent emission has the same origin as the weak bursts (however, see \citealt{Lin2012} and \citealt{Lin2013} who found the opposite using data from \textit{XMM-Newton} and \textit{Swift}).

Analysis of the temporal properties is also an effective way to research the radiation mechanism. \citet{Cheng1996} discovered the short bursts in \object{SGR 1806-20} and the starquakes have similar temporal characteristics: they both have lognormal waiting time distributions as well as have power law energy distributions $dN \propto E^{-1.6\pm0.2}dE$. Thus, they suggested that short bursts in SGRs may be powered by starquakes. More detailed analysis for \object{SGR 1900+14} \citep{Govgucs1999} and \object{SGR 1806-20} \citep{Govgucs2000} confirmed these former discoveries. \citet{Gotz2004} analyzed the spectral evolution of short bursts in \object{SGR 1806-20} using data from \textit{INTEGRAL} and found a negative relationship between hardness ratio and intensity. In subsequent work, \citet{Gotz2006} confirmed this correlation and analyzed the intensity distribution of short bursts. \citet{Nakagawa2007} showed the spectral and temporal properties for \object{SGR 1806-20} and \object{SGR 1900+14} using data from \textit{HETE-2}. \citet{Woods2005} claimed the existence of two classes of bursts in AXPs/SGRs, basing on the existence of extended X-ray tails (tens to hundreds of seconds) and the correlation with pulses for some bursts. To summarize, the analysis of short bursts, whether spectral or temporal, is important to determine the mechanism of magnetar radiation.

An effective way to locate bursts is by Bayesian blocks algorithm. This algorithm was developed by \citet{Scargle1998} and \citet{Scargle2013} to analyze the structures in photon counting data and to detect Gamma-ray Bursts. \citet{Lin2013} first used this algorithm to search for short bursts in SGRs, and they found the technique to be especially helpful in distinguishing dim bursts. They analyzed the morphological properties of the short bursts and fitted the duration distributions with two lognormal functions for \object{SGR 0501+4516}, and then they verified the power law distribution of the fluence. As a Bayesian method, the Bayesian blocks algorithm inevitably has prior parameters to determine. Furthermore, this algorithm has time complexity $ O(n^2) $ \citep{Scargle2013}, so additional work is necessary to reduce the computing time. More details will be shown in Section 3.

In this paper, we analyze the temporal and spectral properties of three AXPs/SGRs: \object{SGR 1806-20}, \object{1E 1048-5937} and \object{SGR 0501+4516}. We locate short bursts using the Bayesian blocks algorithm, and we analyze the spectral and temporal data with the aim to constrain the potential energy origins of bursts. In Section 2, we describe the observations and data reduction. The details of detecting short bursts by using tje Bayesian blocks algorithm are presented in Section 3. In Section 4, we show that the short burst duration distributions, evolution of the flux and the relationship between short bursts and persistent emission. We discussed the accretion model and the magnetar model, basing on our results, in Section 5.

\section{Observations and data reduction}

We chose three sources for our research, including \object{SGR 1806-20}, \object{1E 1048-5937} and \object{SGR 0501+4516}. The objects were observed by \textit{XMM-Newton} space telescope. All these sources have enough observations to study short bursts statistically.

\object{SGR 1806-20} was one of the first SGRs to be discovered \citep{Laros1987}. It is also one of the most energetic sources among all AXPs/SGRs, with an outburst lasting $\sim 10 ~\rm yrs$. In December 2004, it showed the third giant flare \citep{Borkowski2004} ever observed in SGRs, releasing $\sim 10^{46} \rm~ erg$ during $\sim 0.5~ \rm s$ \citep{Mereghetti2008}. \textit{XMM-Newton} monitored the source for about 10 years and observed its entire outburst variation. \citet{Woods2007} showed the evolution of \object{SGR 1806-20} before and after the giant flare. \object{1E 1048-5937} is a bright AXP, and it experienced two outbursts in 2002-2004 and 2007 \citep{Tam2008}. \textit{XMM-Newton} has one observation for each outburst, respectively. \object{SGR 0501+4516} was discovered on 2008-8-22 when it entered its first outburst \citep{Barthelmy2008}. \textit{XMM-Newton} observations of this source began on 2008-8-23, showing an entire outburst decay similar to that of \object{SGR 1806-20} \citep{Camero2014}.

We analyzed the data from the detector PN on board \textit{XMM-Newton}, since it has the highest time resolution in imaging mode among all three soft X-ray detectors. Almost all of the observations are in the imaging modes, only one observation for \object{1E 1048-5937} is in timing mode. We only used the data in imaging mode, to make the results from different observations comparable. We obtained time-tagged events (TTE) data from circular source regions with radii of $30^{''}$ centered on the sources' positions and performed background subtraction using another circle regions with radii of $45^{''}$ aside these sources. Using these TTE data, we located the non-piled-up short bursts by the Bayesian blocks algorithm. The details about the Bayesian blocks algorithm will be discussed in Section 3.

Since the count rates of some bursts are high, the pileup effect must be considered to ensure the precision of burst spectra. Considering the lack of photons for some bursts, two different ways were used to detect pileup for bursts with $>50$ counts and those with $\leq 50$ counts. For the former bursts, results from \textit{epatplot} in \textit{XMM-SAS} were used to determine the presence of pileup. We used the single pattern fractions between the observed data and models given by \textit{XMM-SAS}. The bursts with the fractions of 0.950 to 1.050 were marked as non-piled-up bursts, while others were filtered out of burst data as piled-up bursts. In the normal case (with enough photons), this fraction should not be much larger than 1, since pileup only lowers this fraction. But considering the uncertainty in the statistics of observed patterns, this fraction can be quite large, like $\sim 1.1$. We excluded these bursts with larger fractions, mainly because they deviate from the model significantly, which reflects some anomalies in observed pattern distribution. We counted the photons in single ($N_{\rm s}$), double ($N_{\rm d}$), triple ($N_{\rm t}$) and quadruple ($N_{\rm q}$) patterns for bursts which lack photons. Then we calculated the ratio ($\frac{N_{\rm t}+N_{\rm q}}{N_{\rm s}+N_{\rm d}}$), where $N_{\rm t}+N_{\rm q}$ represents the number of anomalous photons. The burst with ratio larger than 0.1 are regarded as a piled-up burst. The burst detection results without pileup are shown in \reftable{shortBurst}.

\begin{figure}
\centering
\subfigure{
\includegraphics[width=0.3\textwidth,angle=270]{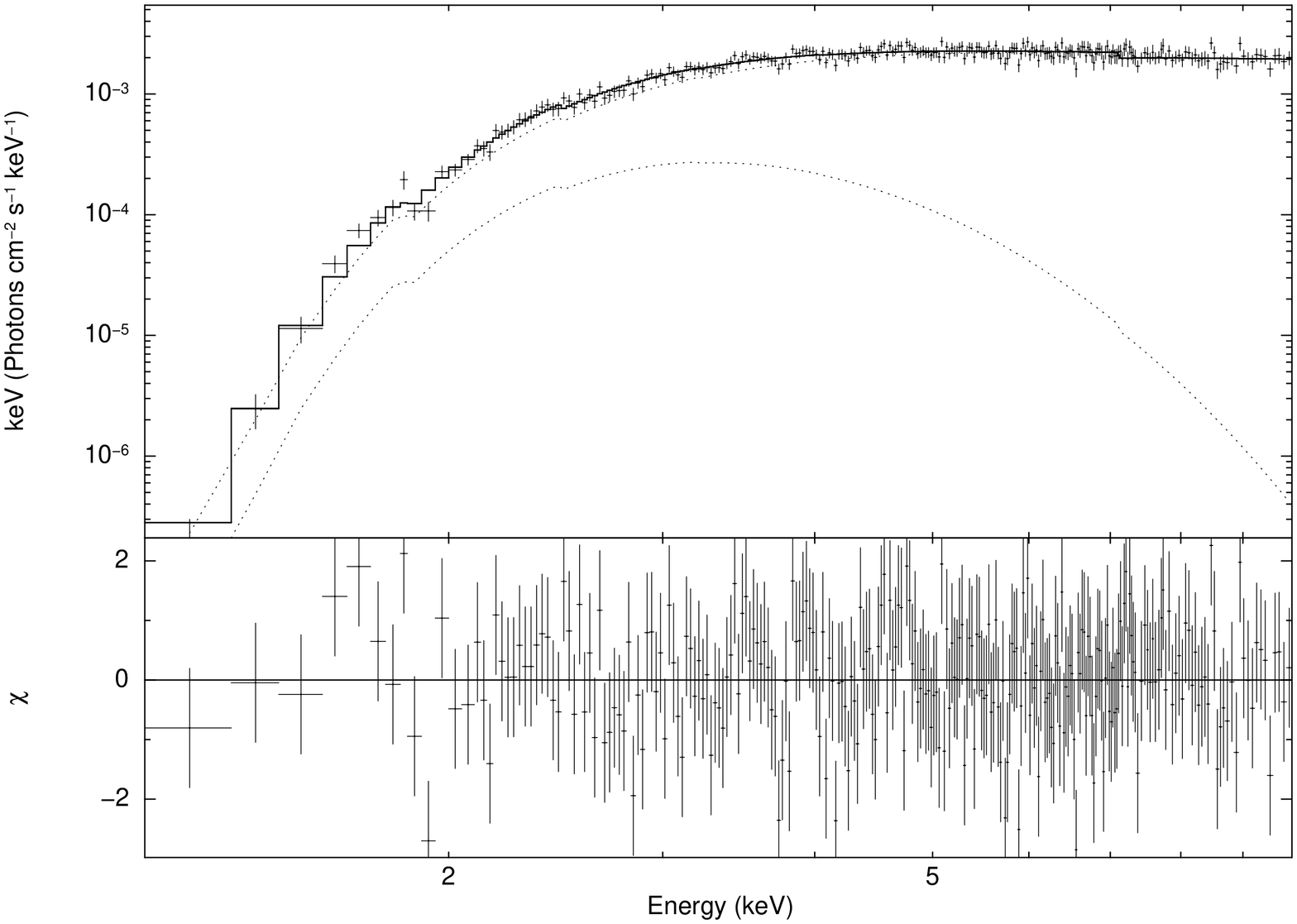}
\hspace{0.05\textwidth}
\includegraphics[width=0.3\textwidth,angle=270]{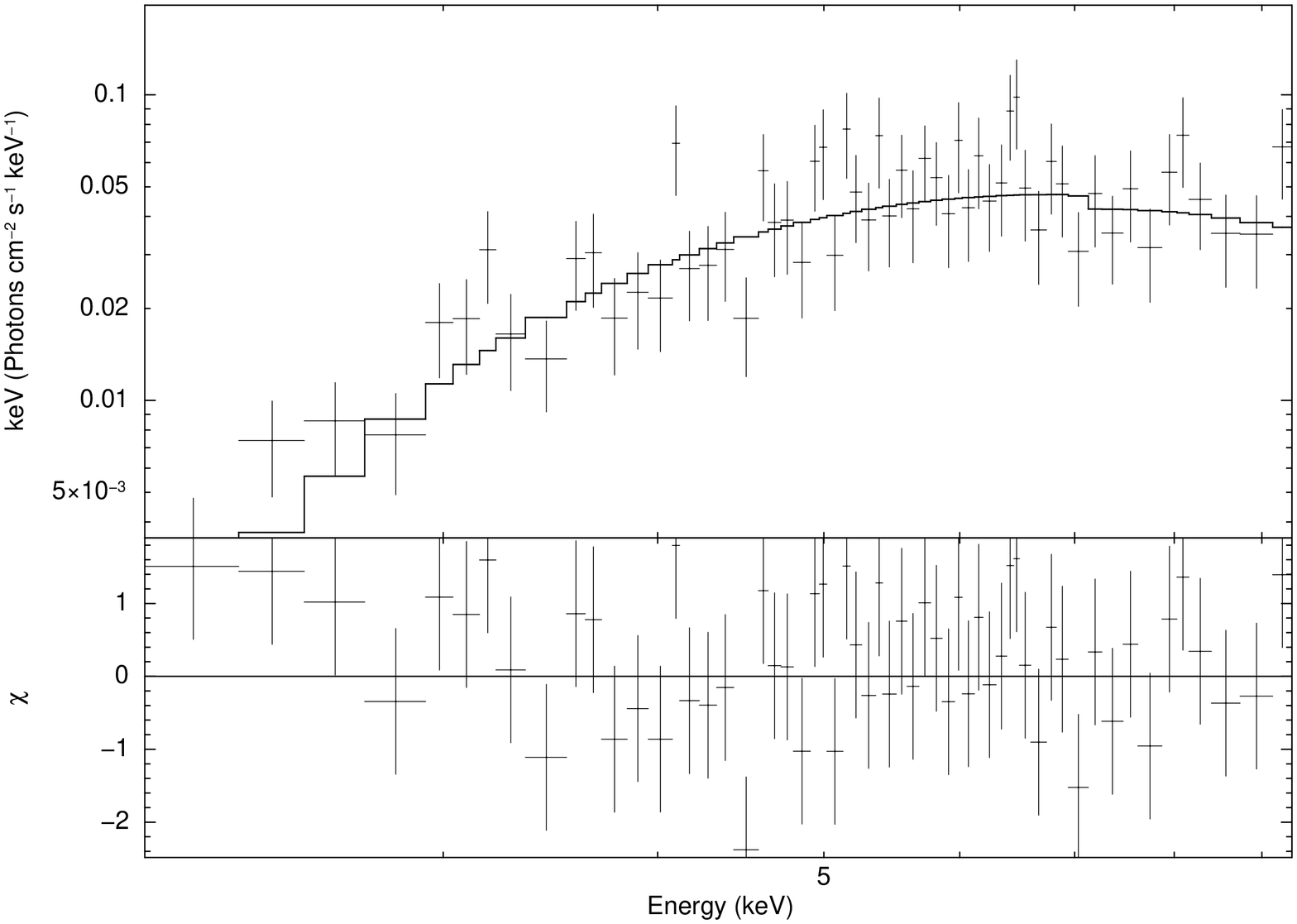}
}
\caption{The unfolded spectra of persistent emission (left panel) and bursts (right panel) from Obs. ID 0164561101 of SGR 1806-20. We utilized absorbed black body plus power law to fit the persistent emission and a single absorbed black body to fit the bursts.}
\label{specurve}
\end{figure}

In order to compute the flux and luminosity, we fitted the persistent spectrum for each observation and burst spectra for the observations with more than 50 burst photons. According to the former works \citep{Fenimore1994, Mereghetti2005, Tiengo2005, Rea2009}, we adopted an absorbed black body plus power law (\textbf{phabs(bbodyrad+powerlaw)} in Xspec) for the persistent emissions and a single absorbed black body for bursts. The BB+PL model is a simplified phenomenological model for the persistent emission of AXPs/SGRs which assumes the emission is the sum of a blackbody from a hot spot on the stellar surface and a non-thermal component enhanced by resonant cyclotron scattering. In some observations of our targets, a single blackbody component cannot account for the burst spectra, so we added a second blackbody component when modeling these spectra. The photoelectric absorption parameter $N_{\rm H}$ was fixed using the value from magnetar catalog \citep{Olausen2014} shown in \reftable{param}, except for in the modeling of the \object{SGR 0501+4516} data shown in \reftable{spec0501}. In the spectra fitting, the background spectra are different for the persistent spectra and the burst spectra. For persistent spectra, the background spectra are extracted from the background sky described above. For the burst spectra, we employed the persistent spectra as their background spectra. Examples of the persistent and the burst spectra are shown in \reffig{specurve}, respectively. After the spectral fitting, the \textit{cflux} in Xspec was used to calculate the unabsorbed flux in the energy band of $1.0~-~10.0~\rm keV$ and the error with a 90\% confidence range. To obtain the luminosity in the same band, we used the distances in \reftable{param}. The results are listed in  \reftable{shortBurst}~ -~ \reftable{spec0501}.

Because high-rated piled-up bursts have been excluded, the obtained burst fluxes should be lower than the real ones for observations containing piled-up bursts. Besides pileup effect, high count rates might cause telemetry saturation. Telemetry saturation had two origins: short bursts and soft proton flares (SPFs). SPFs have been filtered out using the background light curves, so the telemetry saturation related to SPFs only makes observation time equivalently shorter. However, the telemetry saturation caused by short bursts would decrease burst fluxes because of the zero-rated intervals (gaps) when the detector switched into timing mode. Combining these two effects together, it is clear that we obtained lower limits on the burst fluxes and unbiased persistent fluxes.

\section{Details of locating short bursts}
In the Bayesian blocks algorithm, the optimal block partition is obtained by maximizing the likelihood function ($L$) \citep{Scargle2013},
\begin{equation}
{\rm ln} L^{(k)}=N^{(k)}{\rm ln} \lambda^{(k)}-\lambda^{(k)}T^{(k)},
\end{equation}
where $N^{(k)},~\lambda^{(k)},~T^{(k)}$ are the total number of photons, the expected count rate, and the duration of the block $k$, respectively. The algorithm depends on the prior blocks distribution index ($ncp\_prior$) and false positive probability $p_0$. We utilized the equation (21) in \citet{Scargle2013} ($ncp\_prior=4-{\rm{log}} (73.53p_0N^{-0.478})$) to determine $ncp\_prior$ and chose 0.05 as the default value of $p_0$.

Although the time complexity has been reduced to $ O(n^2) $ \citep{Scargle2013}, it is still an unacceptable computing time if we apply the Bayesian blocks algorithm to the entire observation. To reduce the time complexity again, we divided the entire observation into several segments of equal length. For different sources, we defined the length as the mean photon counts received in $50\rm~ s$, but at least 100 counts.

\begin{figure}
\centering
\subfigure{
\includegraphics[height=0.28\textheight]{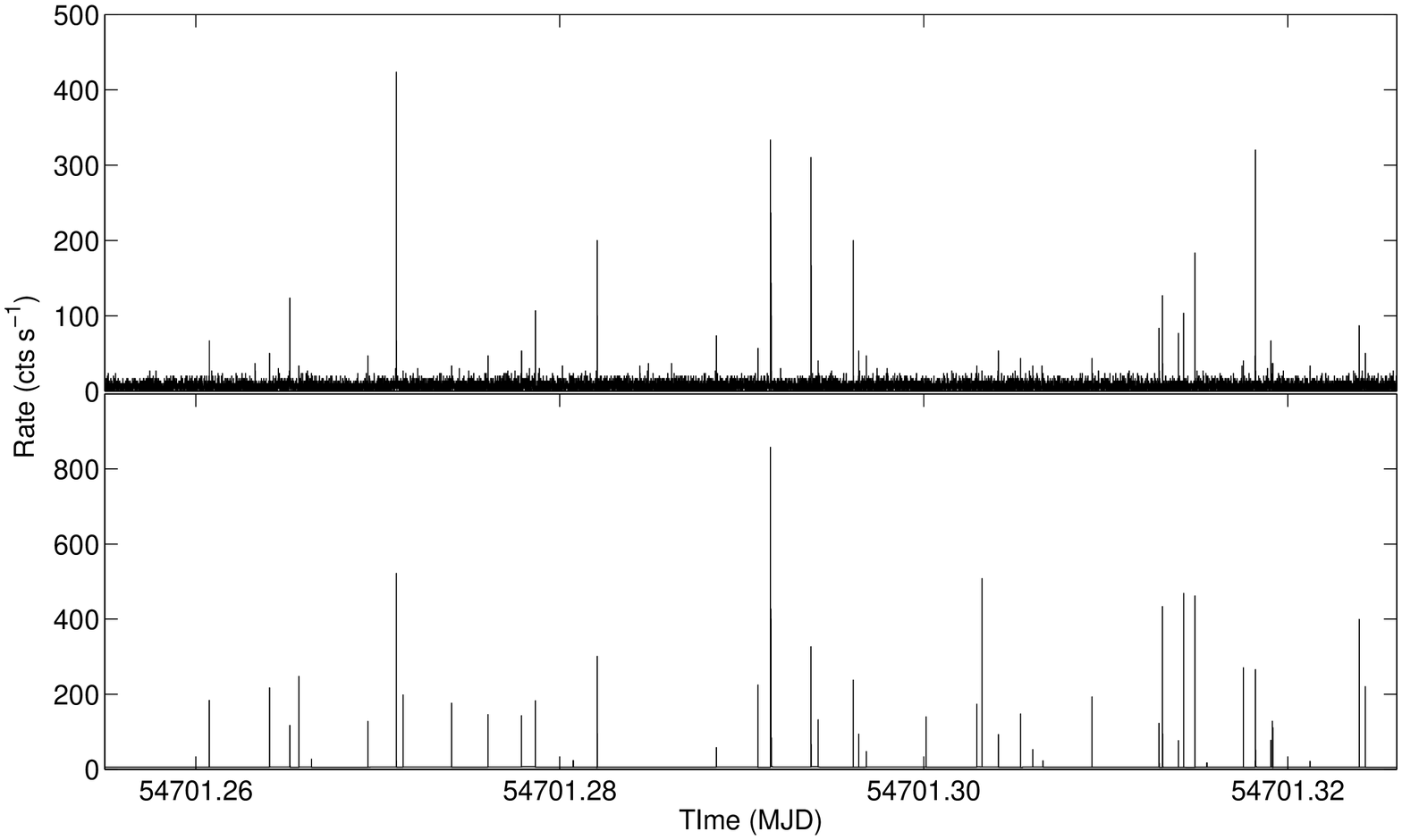}
\includegraphics[height=0.28\textheight]{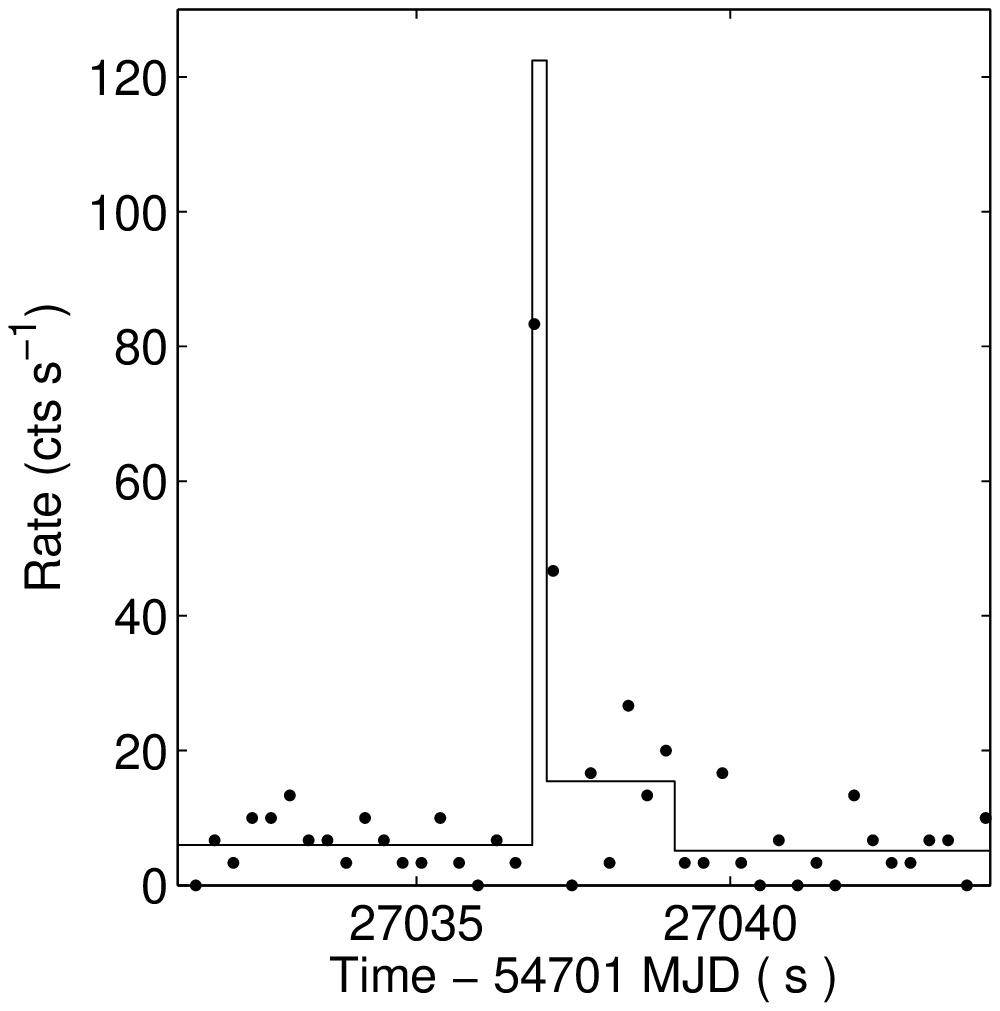}
}
\caption{The data are from parts of SGR 0501+4516's first observation (Obs. ID 0560191501). Left panel: the photon count rates with $0.3~\rm s$ bin (top) and the short burst results (bottom) of Bayesian blocks algorithm.  The duration is about $ 6000~\rm s$. Right panel: one detected short burst in $10 ~ \rm s$ observation. The dots are the $0.3~\rm s$ bin data, while the solid line is the result of Bayesian block algorithm.}
\label{example}
\end{figure}

To decrease the effect of the different separations, we set two rounds of detection. The first round started from the beginning of the observation, and the second round was a half separation postponed relative to the first one. For each round, we applied the Bayesian blocks algorithm to each segment, and we obtained the raw change points. If one change point was also a discontinuity points we set in the separations, we used the two segments before and after it to determine whether it was a change point or not. We merged the two round results together and calculated the final rate for each block. We defined the blocks longer than double periods of the source as the background blocks, and we amalgamated the background blocks to estimate the background count rate level. The gaps in data set were regarded as blocks with zero rates by the Bayesian blocks algorithm, and we excluded them out from the background rate computation. Subsequently, we tagged the blocks with rates higher than that of the background as short burst blocks. An example of the results from the Bayesian blocks algorithm is shown in \reffig{example}. All background blocks for each observation were collected to constitute the persistent data, while the burst blocks were collected as the burst data.

\section{Results}
\subsection{The temporal properties}

\begin{figure}
\centering
\subfigure{
\includegraphics[width=0.33\textwidth]{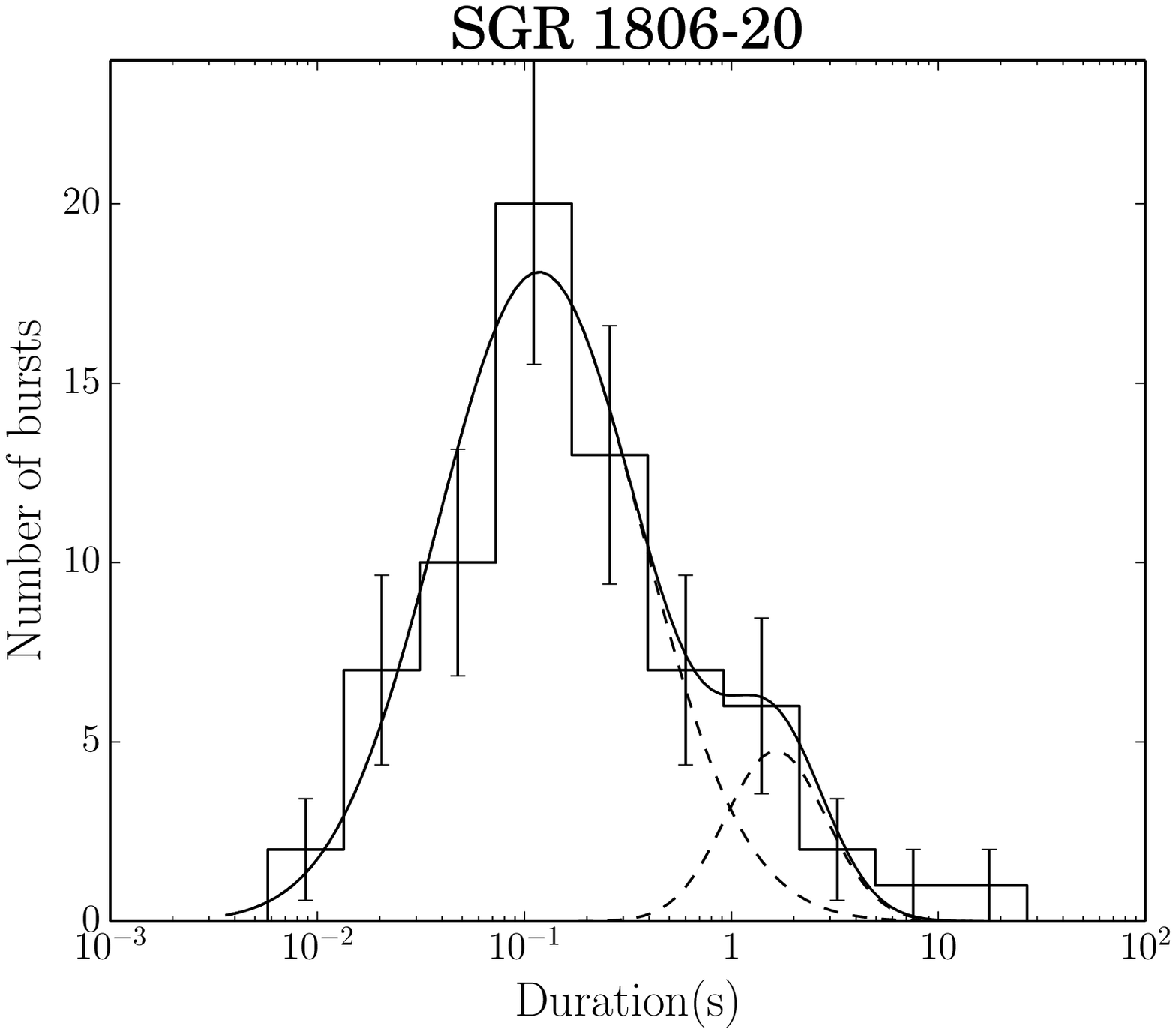}
\includegraphics[width=0.33\textwidth]{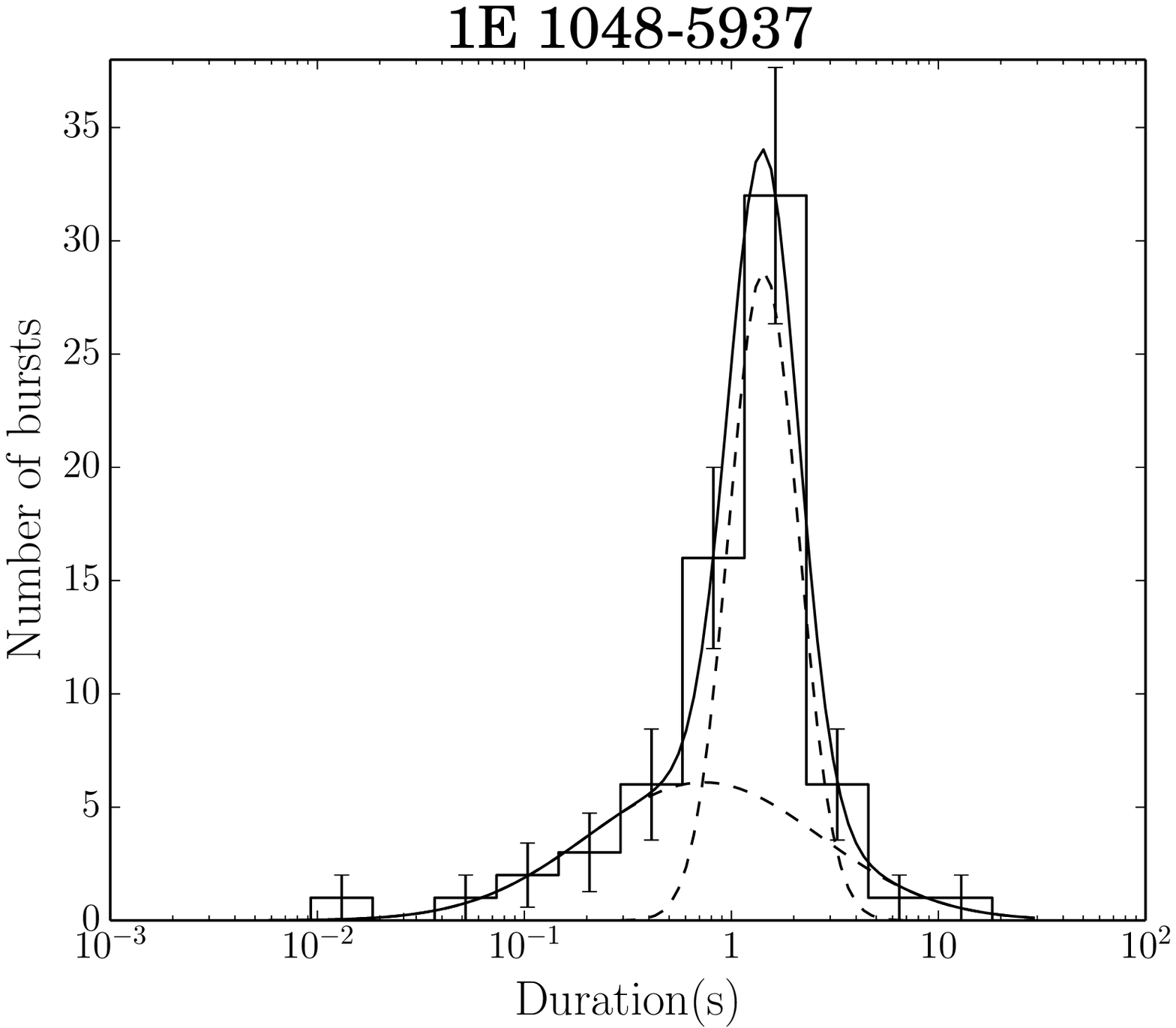}
\includegraphics[width=0.33\textwidth]{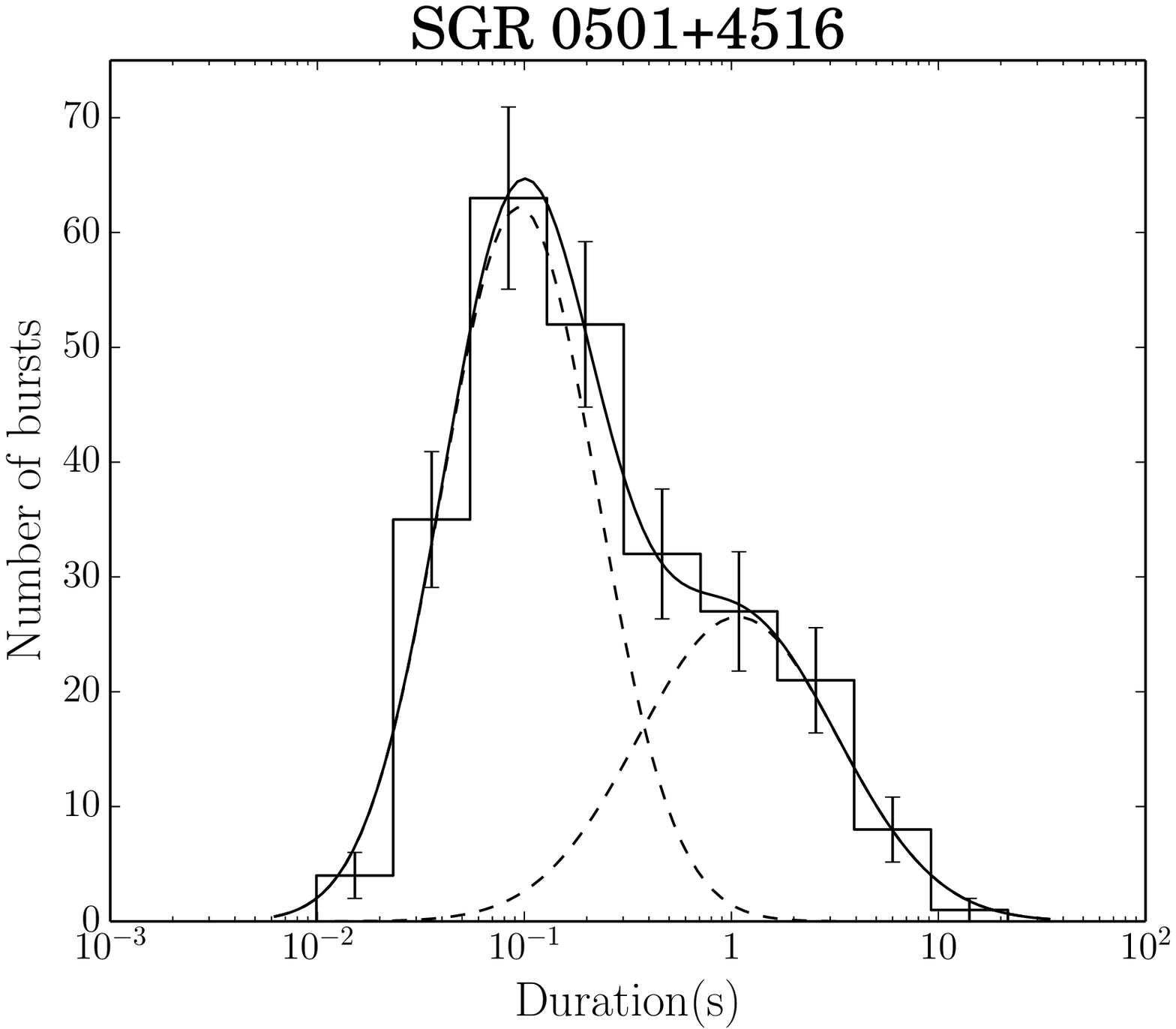}}
\caption{The duration distributions for SGR 1806-20, 1E 1048-5937 and SGR 0501+4516 from left to right, respectively. The solid lines are the best fit of the histogram with a sum of two lognormal functions, while the dashed lines are the two components respectively.}
\label{duration}
\end{figure}

After locating the short bursts, we made the duration distributions for \object{SGR 1806-20}, \object{1E 1048-5937} and \object{SGR 0501+4516}. We found that each distribution can be fitted by the sum of two lognormal functions. All of the results are shown in \reffig{duration}. For \object{SGR 1806-20}, the two components are $ \tau_1 = {117 \pm 3}~\rm ms $ with standard deviation $ \sigma_1 = 0.49 \pm 0.08 $ and $ \tau_2 ={1.6_{-0.2}^{+1.0}}~\rm s $ with $ \sigma_2 = 0.24 \pm 0.19$.  For \object{1E 1048-5937}, the two components are $ \tau_1 = {0.73_{-0.12}^{+0.14}}~ \rm s $ with $ \sigma_1 = 0.56 \pm 0.04 $ and $ \tau_2 = {1.42 \pm 0.03}~ \rm s $ with $ \sigma_2 = 0.17 \pm 0.01$. For \object{SGR 0501+4516}, the two components are $ \tau_1 = {94 \pm 7}~ \rm ms $ with $ \sigma_1 = 0.37 \pm 0.02 $ and $ \tau_2 = {1.08_{-0.21}^{+0.26}}~ \rm s $ with $ \sigma_2 = 0.48 \pm 0.07$. The $ {\chi}^2/ d.o.f$ are $3.76/5 $, $1.66/6 $ and $2.41/4 $, respectively.

\begin{figure}
\centering
\plottwo{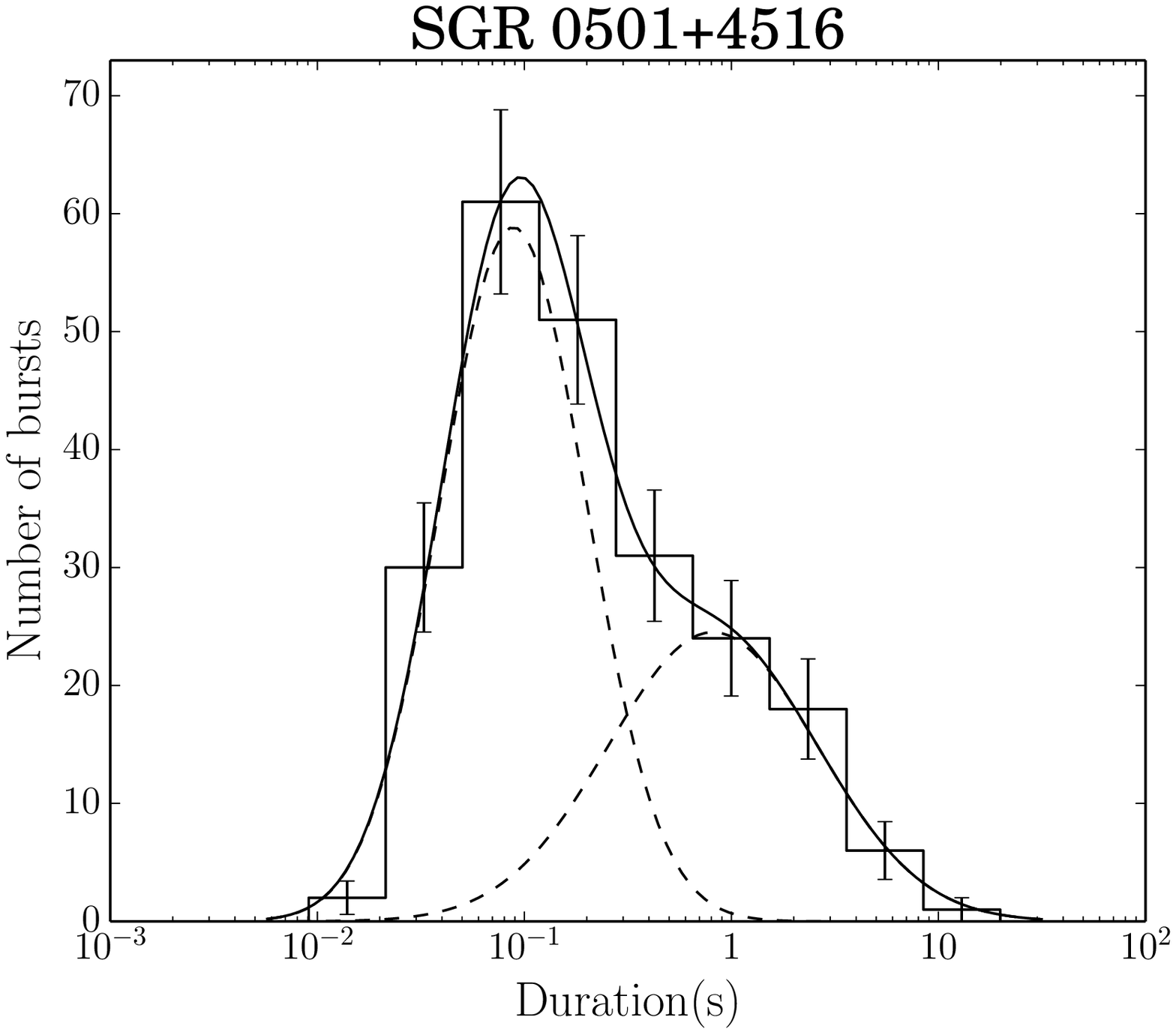}{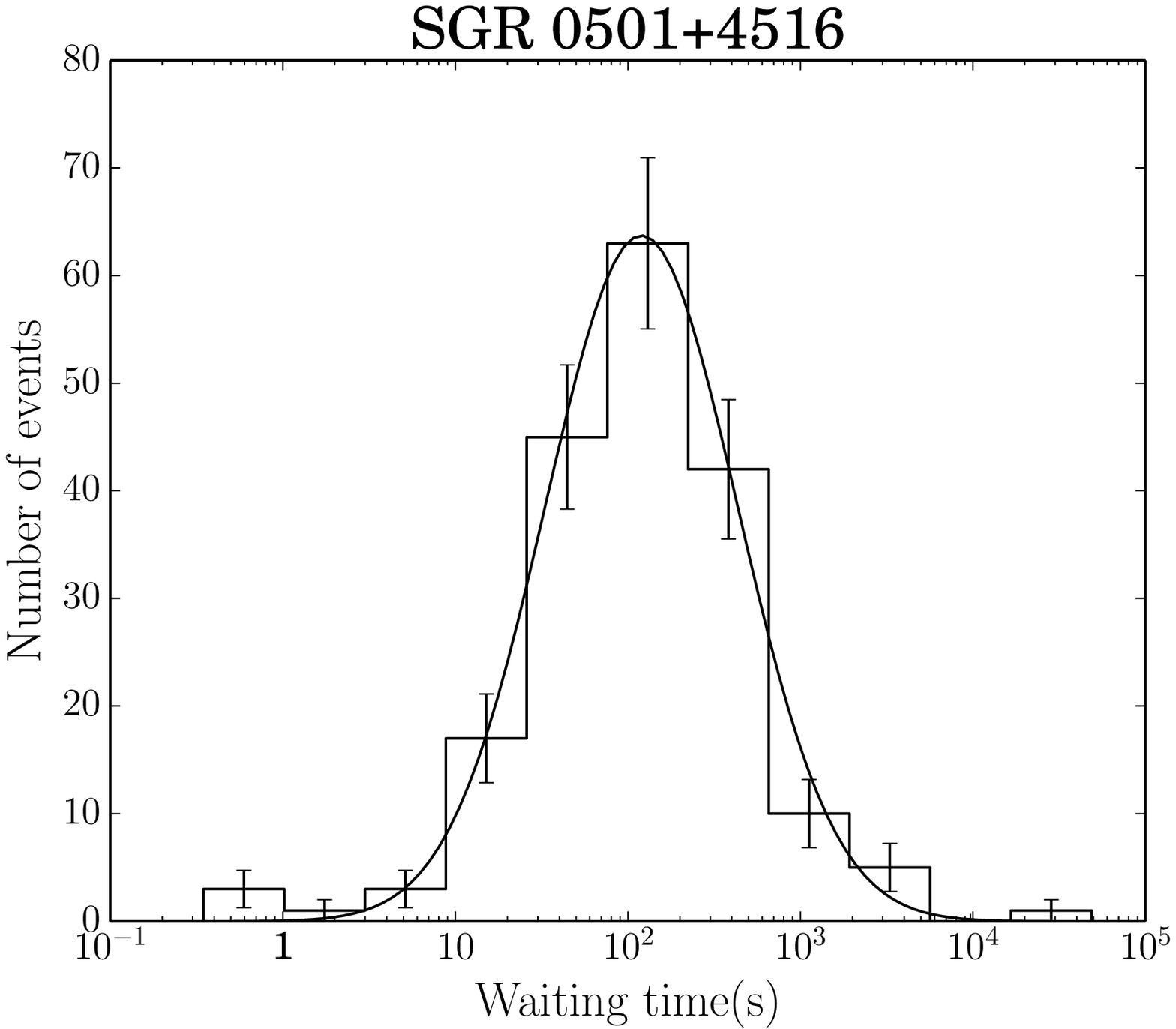}
\caption{Right panel shows the duration distribution of the first observation from SGR 0501+4516 to make a comparison with the results from \citet{Lin2013}. The solid line represents the best fit by sum of two lognormal functions, while the dashed lines represent the two components respectively. Left panal shows the waiting time distribution for all observations from SGR 0501+4516. The solid line is the best fit using lognormal function.}
\label{0501}
\end{figure}

Especially for the first observation of \object{SGR 0501+4516} (Obs. ID 0560191501), \citet{Lin2013} analyzed the duration of the dim short bursts. To make a comparison with their results, we made the duration distribution for this observation particularly, shown in \reffig{0501}. Our results showed that $\chi^2/d.o.f. = 5.86/6$ and two lognormal components of $\tau_1 = {92 \pm 5}~\rm ms$ with $\sigma_1 = 0.38\pm 0.02$, as well as $\tau_2 = {1060_{-128}^{+145}}~ \rm ms$ with $\sigma_2 = 0.35 \pm 0.05$, while \citet{Lin2013} showed that $\chi^2/d.o.f. = 3.65/5$ and two lognormal components of $\tau_1 = {85 \pm 8}~\rm ms$ with $\sigma_1 = 0.36\pm 0.03$ and $\tau_2 = {1028_{-181}^{+220}}~ \rm ms$ with $\sigma_2 = 0.32 \pm 0.05$. Since the Bayesian block algorithm has some prior parameters, we get a different bursts sample compared to \citet{Lin2013}. However, considering the uncertainties, our results are consistent with \citet{Lin2013}. 

\begin{figure}
\centering
\subfigure{
 \includegraphics[width=0.3\textwidth,angle=270]{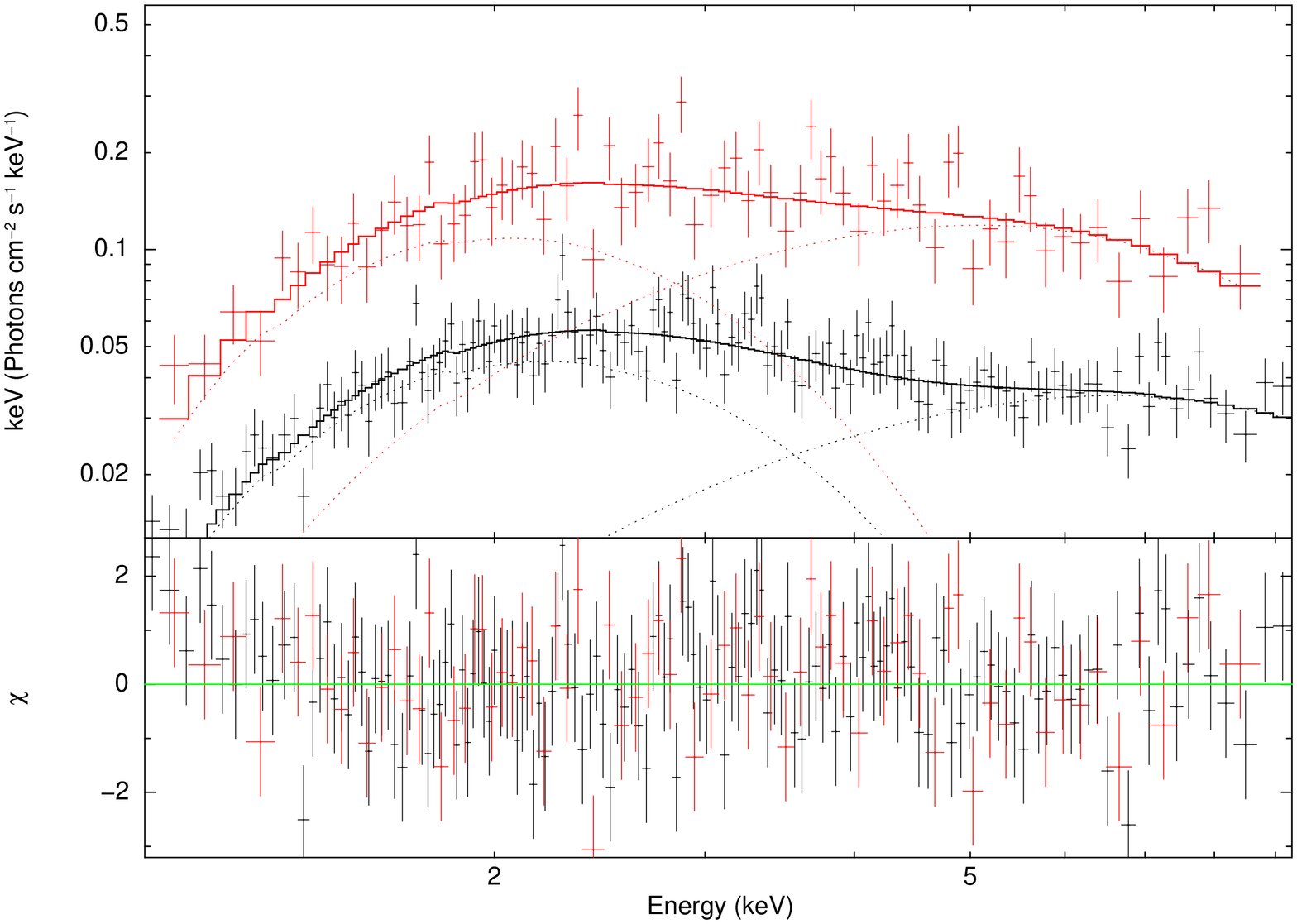}
 \includegraphics[width=0.3\textwidth,angle=270]{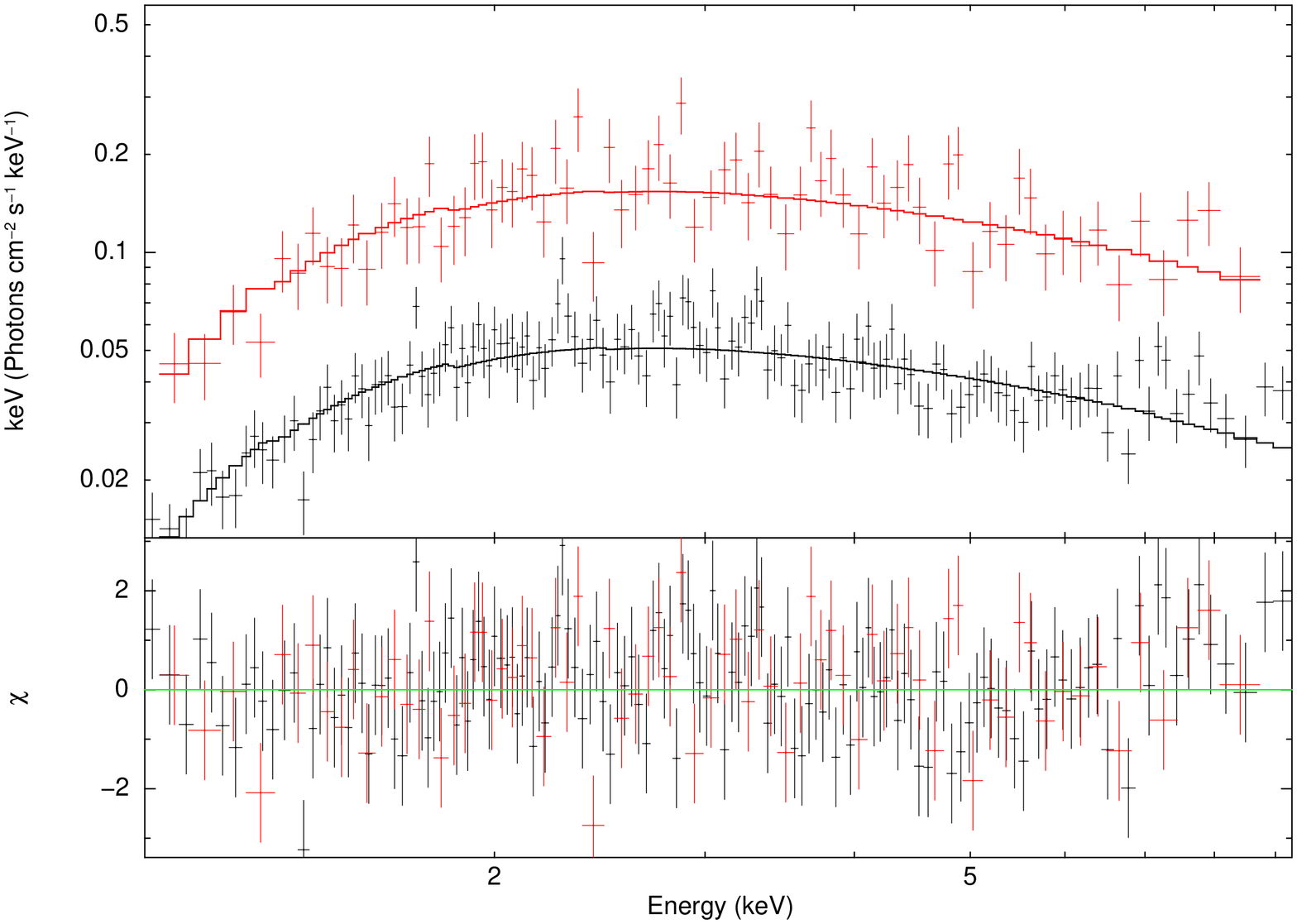}
}
 \caption{The spectra of short time scale bursts ($< 0.2 ~\rm{s}$) with high normalization and long time scale bursts ($\geq 0.2~\rm{s}$) with low normalization for SGR 0501+4516, respectively. Both of them can be well fitted by two black body components (left), or by OTTB model (right).}
 \label{svl}
\end{figure}

We noticed that the bursts are mainly dominated by the short time scale bursts ($\sim 0.1~\rm s$) in \object{SGR 1806-20} and \object{SGR 0501+4516}, while the long time scale bursts ($\sim 1~\rm s$) are in the majority in \object{1E 1048-5937}. To investigate the spectral properties of short time scale bursts and long time scale bursts, we divided the bursts into two subclasses at 0.2 s for \object{SGR 0501+4516}, since \object{SGR 0501+4516} has enough burst photons for spectral fitting and $0.2~ \rm s$ is the approximate intersecting point of these two components. Both spectra can be well fitted by the sum of two black body components (BB+BB) or the optically thin thermal bremsstrahlung (OTTB) model, shown in \reffig{svl}. In BB+BB model, the two components are $0.59 \pm 0.04 \rm~ keV$ and $2.31_{-0.31}^{+0.47} \rm~ keV$ with reduced $\chi^2 = 1.01 (144)$ for longer bursts, and $0.52_{-0.09}^{+0.10} \rm~ keV $ and $1.73_{-0.28}^{+0.55} \rm~ keV $ with reduced $\chi^2 = 1.13 (71)$ for shorter bursts. The low energy bands ($< 1.3~\rm keV$) of spectra are a little higher than the model, but the two reduced $\chi^2$ show that these results are still acceptable. In OTTB model, the plasma temperatures are $15.2_{-2.7}^{+3.1} \rm~ keV $ with reduced $\chi^2 = 1.01 (146)$ and $16.4_{-4.4}^{+6.9} \rm~ keV $ with reduced $\chi^2 = 1.06 (73)$ for longer and shorter bursts, respectively. Considering the uncertainty of characteristic temperatures, the main difference is the normalization, which is the emission areas for BB+BB model or the densities of plasma for OTTB. Our results show that the spectrum of shorter bursts have larger normalization in both models.

Waiting time is the interval between two adjacent short bursts. Here, we only considered the waiting time between two adjacent non-piled-up bursts. For these three sources, the waiting times range from several seconds to several hours. Nevertheless, the obtained waiting times are just the phenomenological ones, since we don't have effective means to determine whether a burst is single burst or multi-peaked burst. Using these raw waiting times, we found the waiting time distribution of \object{SGR 0501+4516} could be described as a lognormal function, with $\mu = 119.2^{+4.9}_{-4.7}~ \rm s$, $\sigma = 0.56 \pm 0.01$ and $\chi^2 / d.o.f = 7.61/9$, which is shown in \reffig{0501}. However, for \object{SGR 1806-20} and \object{1E 1048-5937}, the lognormal distribution showed a strange bump in the short time scale. Considering the relationship between the burst strength and the waiting time, our results show that there is no apparent correlation, which is consistent with \citet{Govgucs1999} and \citet{Govgucs2000}.

\subsection{The flux evolution}

\begin{figure}
\centering
\epsscale{0.8}
\plottwo{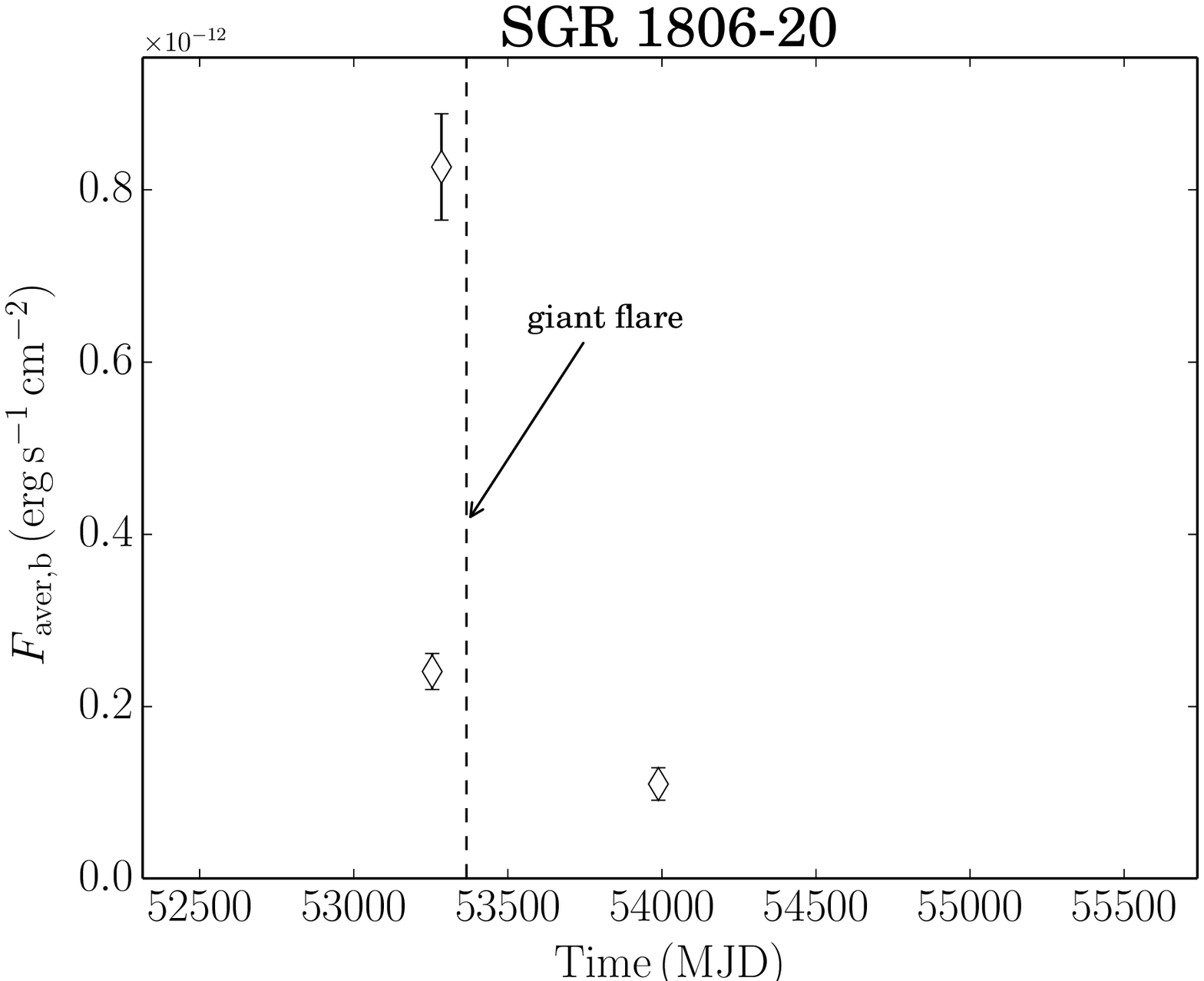}{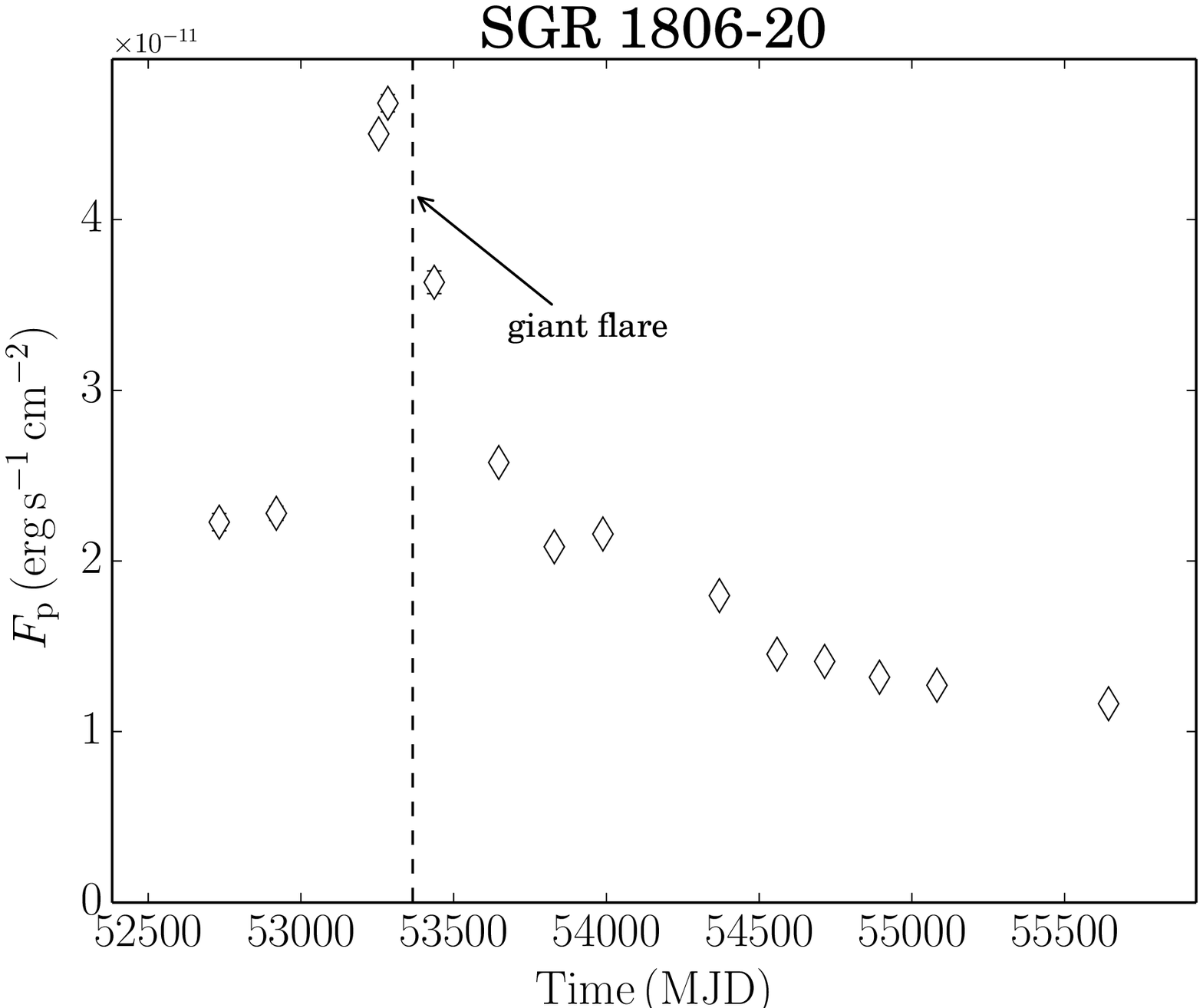}
\plottwo{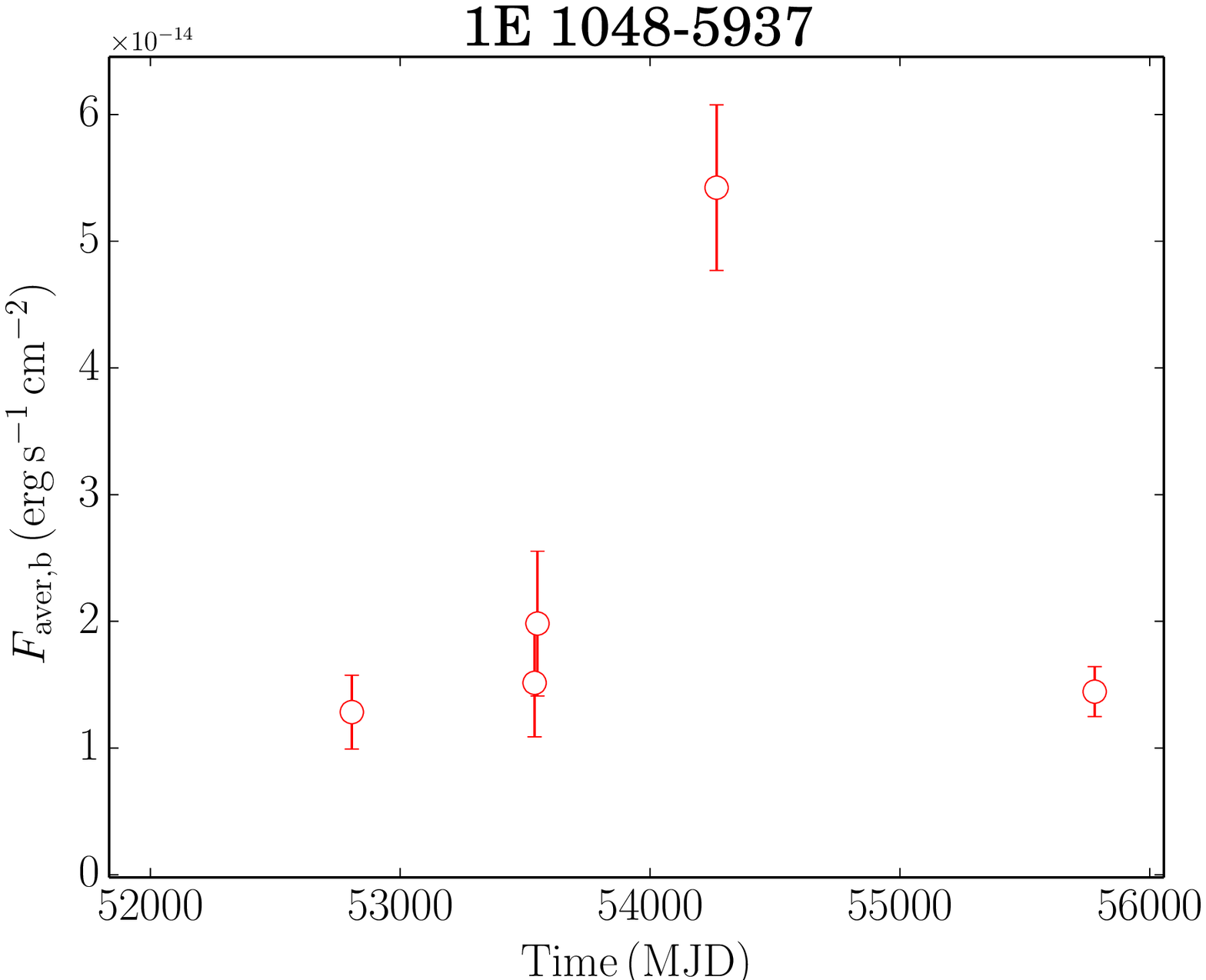}{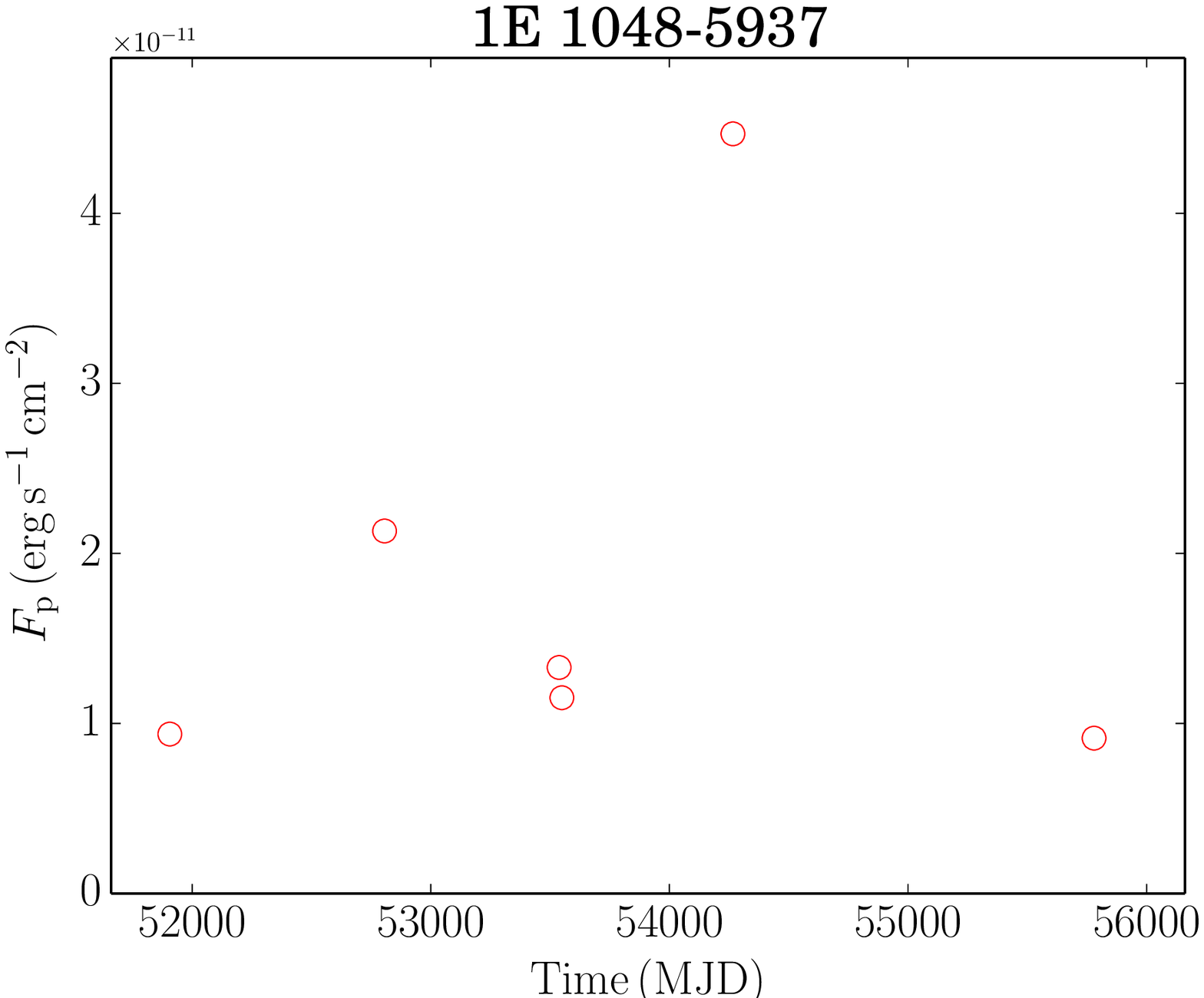}
\plottwo{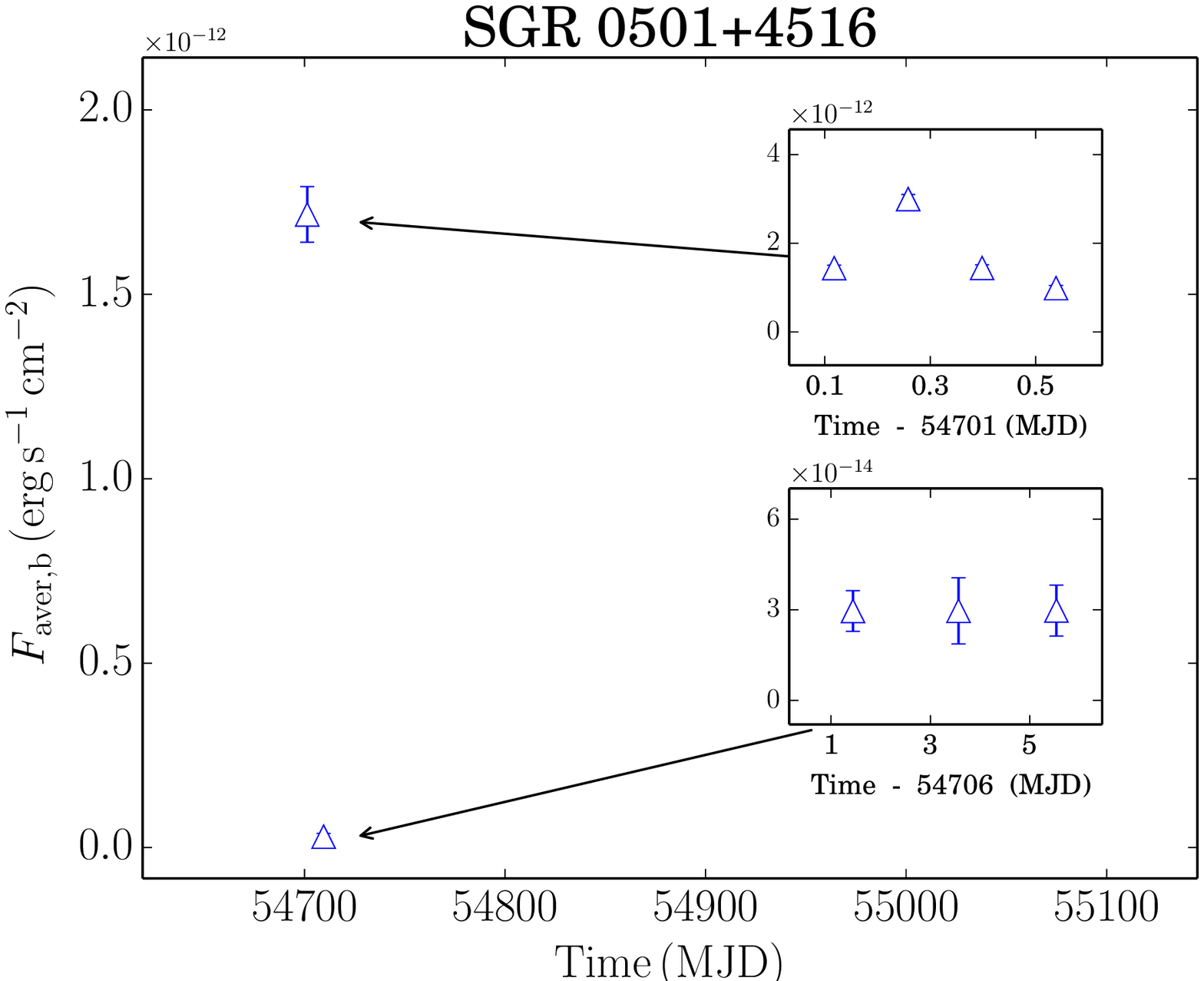}{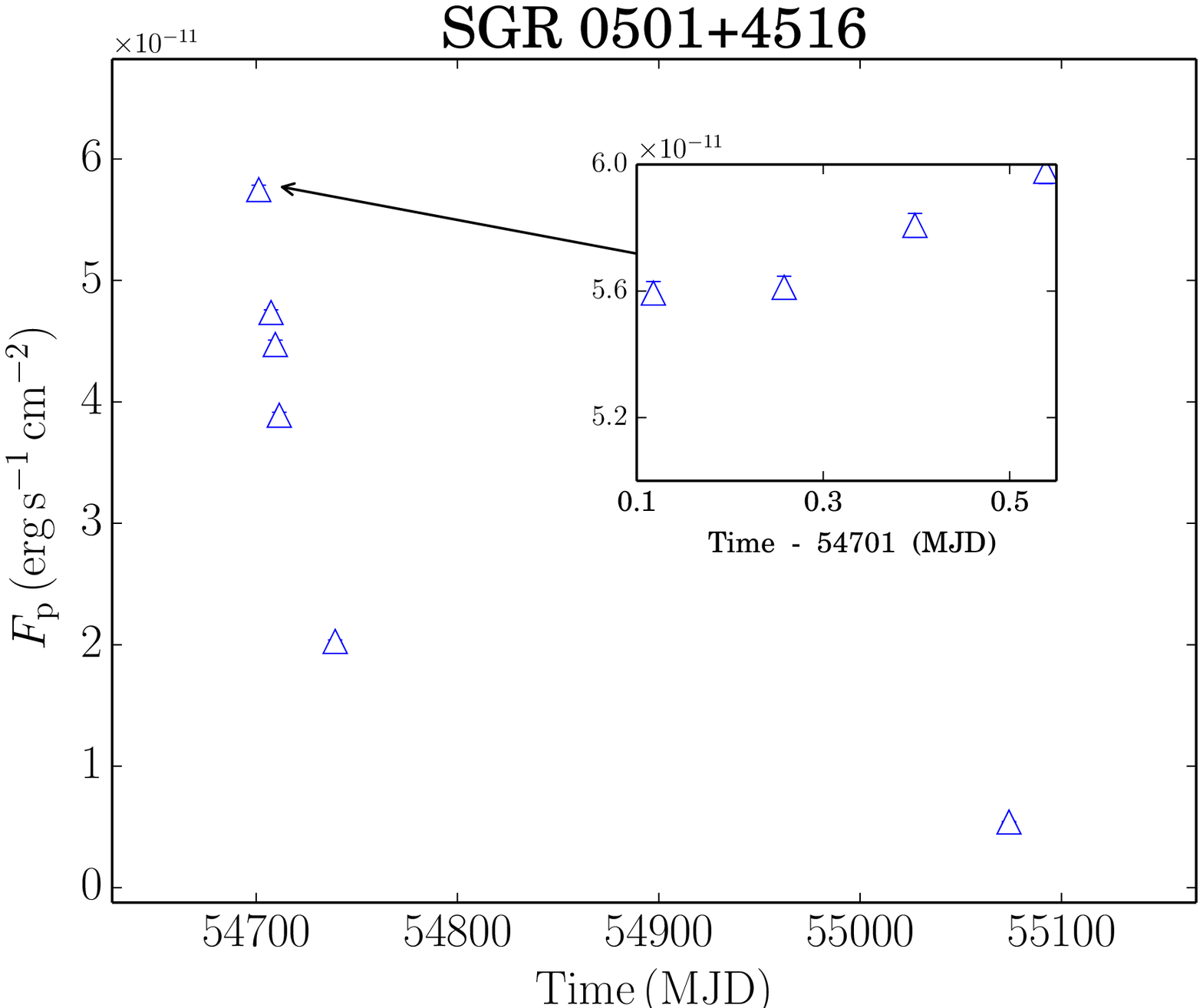}
\caption{The left panels display the persistent flux evolution. The black diamonds, the red circles, the blue triangles represent the persistent flux of SGR 1806-20, 1E 1048-5937 and SGR 0501+4516, respectively (same as \reffig{stat}). The right panels show the average burst flux, only for those observations have more than 50 burst photons, with the same symbols as left panels. Particularly for the SGR 0501+4516, some points are too near to be distinguished, so we calculated the mean value of them and showed the variations in the inside panels.}
\label{lightCurve}
\end{figure}

Based on the spectral fitting results, we plotted the persistent flux ($F_{\rm p}$) and the average burst flux ($F_{\rm aver,b}$) evolution for \object{SGR 1806-20}, \object{1E 1048-5937} and \object{SGR 0501+4516} in \reffig{lightCurve}. The average burst flux denotes that the total burst fluence is averaged into an entire observation, so it represents the strength of burst energy released during each observation.

\subsubsection{SGR 1806-20}

\object{SGR 1806-20} is the most interesting source among all three sources because of its giant flare as we mentioned in Section 2. We noticed that the persistent flux experienced a decay, while the average burst flux showed fluctuation due to the giant flare. After the giant flare, the burst flux became much lower than earlier epoch within three months. During the next two years, it remained at its low burst rate condition until the late in 2006. After a weak peak around MJD 54000, the burst flux entered into another decay stage. Mainly using the data from \textit{RXTE}, \citet{Woods2007} showed the evolution of the frequency, frequency derivative and the burst number per $20$ days. However, their data did not cover the burst peak around MJD 54000.

\subsubsection{SGR 0501+4516}

The persistent flux of \object{SGR 0501+4516} showed a decay \citep{Camero2014}, while the average burst flux showed a steeper drop to the bottom. We separated the first observation into four segments. The persistent flux did not change significantly, while the average burst flux had an apparent peak at the second segment.

\subsubsection{1E 1048-5937}

The flux evolution of this source did not show a decay stage as \object{SGR 1806-20} and \object{SGR 0501+4516}, because \textit{XMM-Newton} did not have enough observations. \object{1E 1048-5937} have experienced two outbursts in 2002-2004 and 2007 \citep{Tam2008}, so the flux variance showed a comparison between the outburst and the quiescent period.

\subsection{Short bursts versus persistent emission}

\begin{figure}
\centering
\plottwo{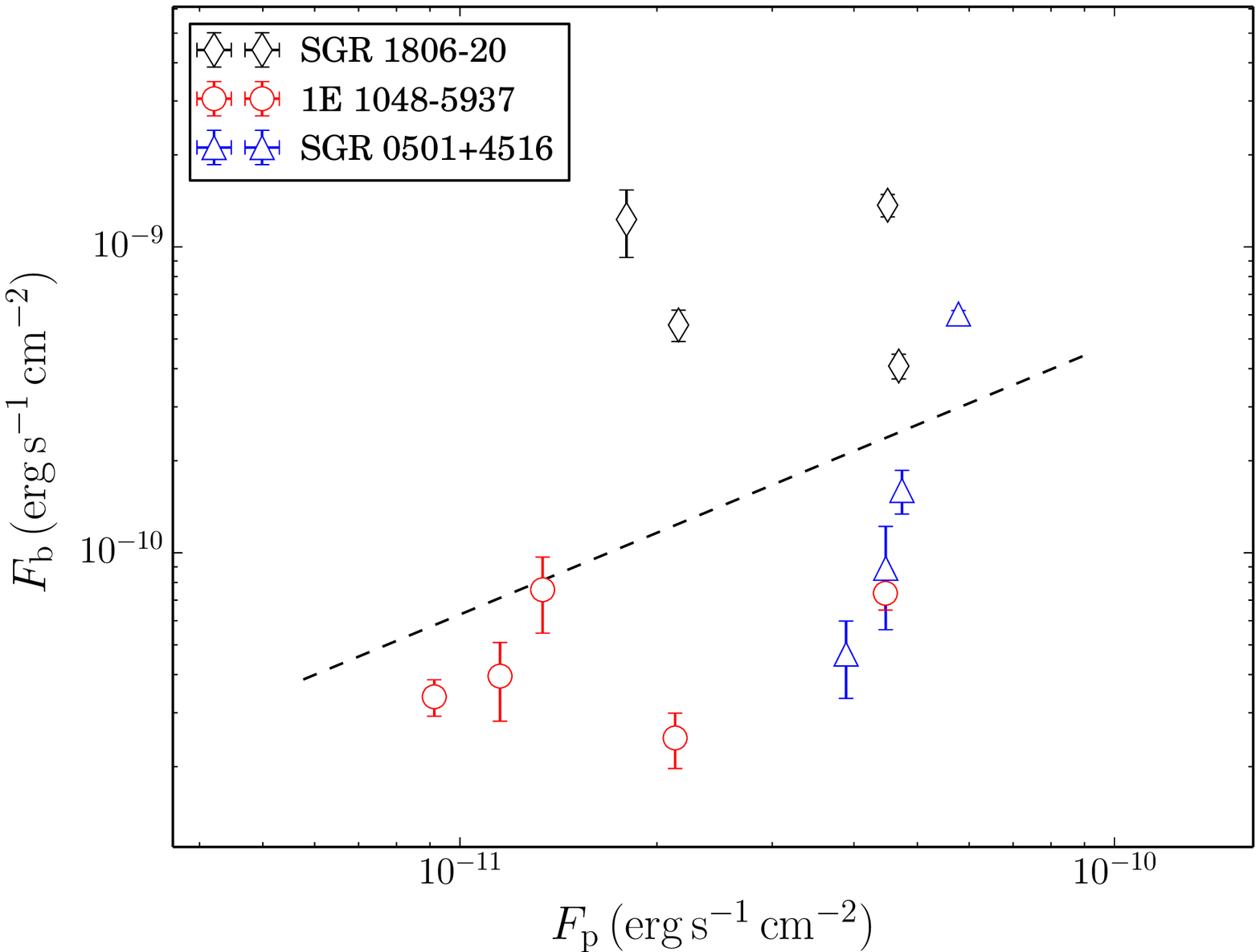}{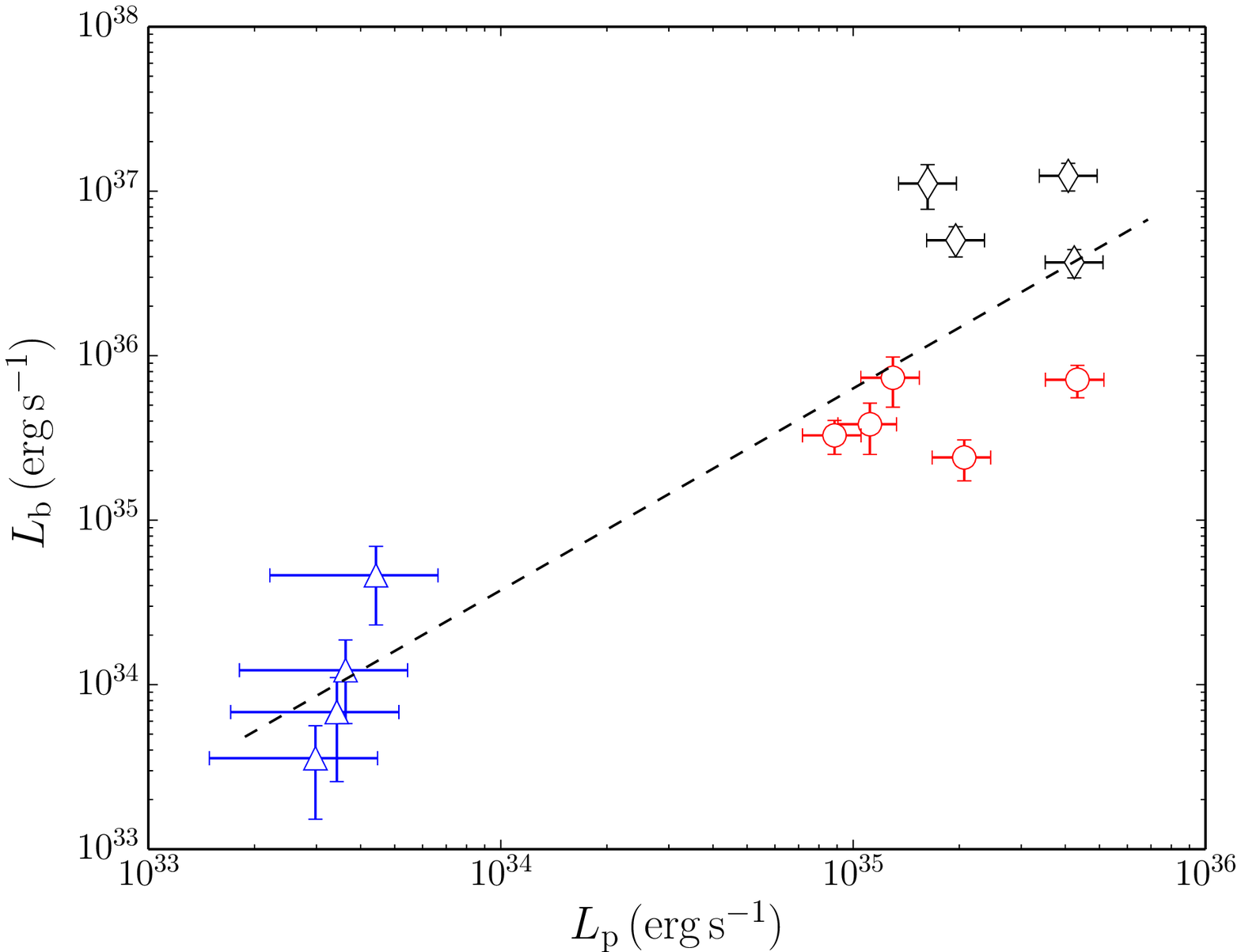}
\caption{The black diamond, the red circle and the blue triangle represent SGR 1806-20, 1E 1048-5937 and SGR 0501+4516 respectively. The dashed lines are the fitted power laws described in Section 4.3.}
\label{stat}
\end{figure}

We also analyzed the relationships between short bursts and the persistent emission of these three sources. We adopted the flux and luminosity to estimate the strength of the short bursts and persistent emission simultaneously. In \reffig{stat}, we fitted $F_{\rm b}$ and ${F_{\rm p}}$ using a power law,
\begin{equation}
F_{\rm b} \propto {F_{\rm p}}^{\gamma},
\end{equation}
with an index of $\gamma = 0.89\pm0.62$ and Pearson correlation coefficient $\rho = 0.40$.  The fit result shows that there is a marginal positive correlation between the burst flux and the persistent flux. We also fitted the burst luminosity and the persistent luminosity using a power law with ${\gamma} = 1.22 \pm 0.18$ and $ \rho = 0.90 $ in \reffig{stat}.  This correlation is more intrinsic than the flux one, which implies that there is a tight relationship between the short bursts and the persistent radiation.

\begin{figure}
\centering
\plottwo{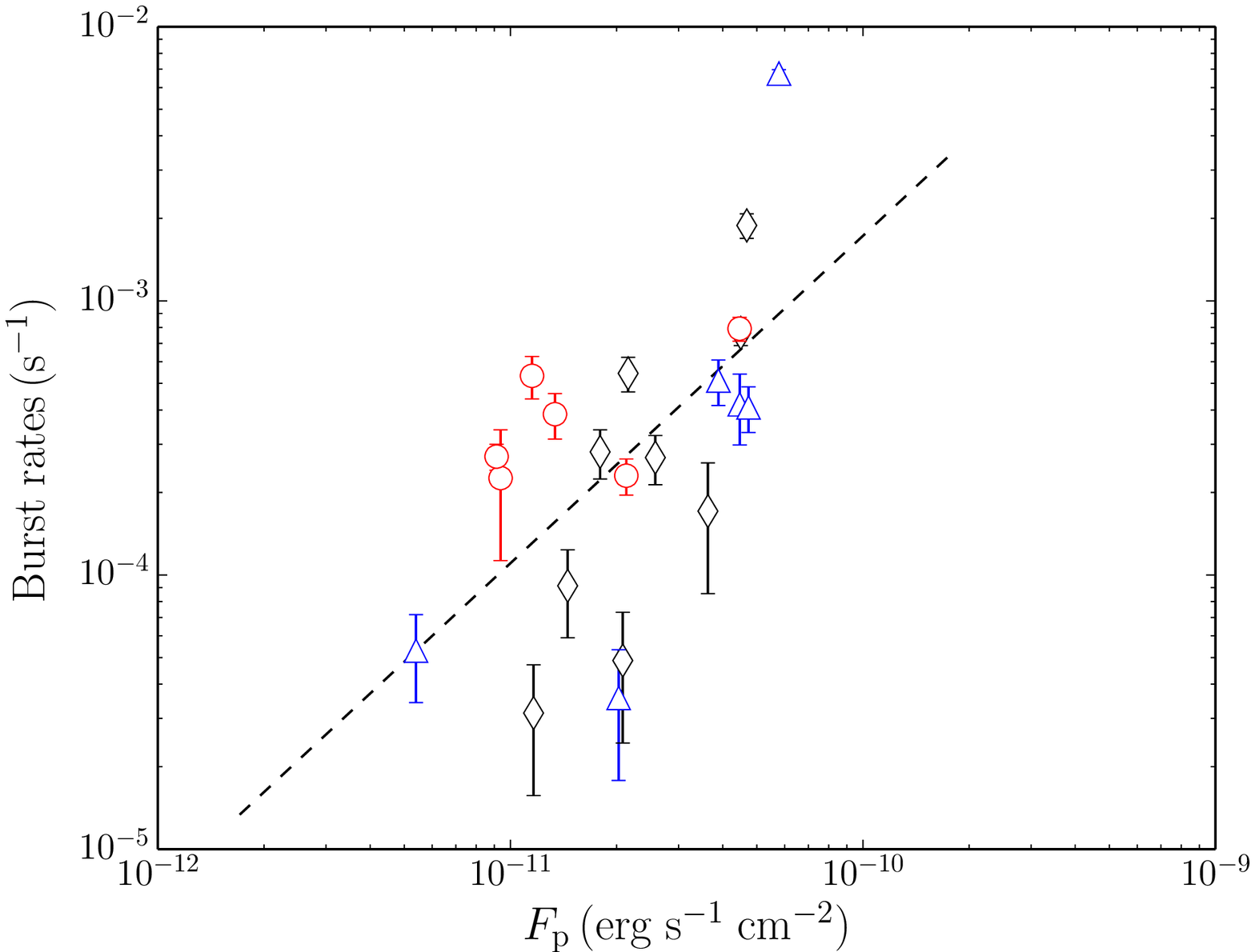}{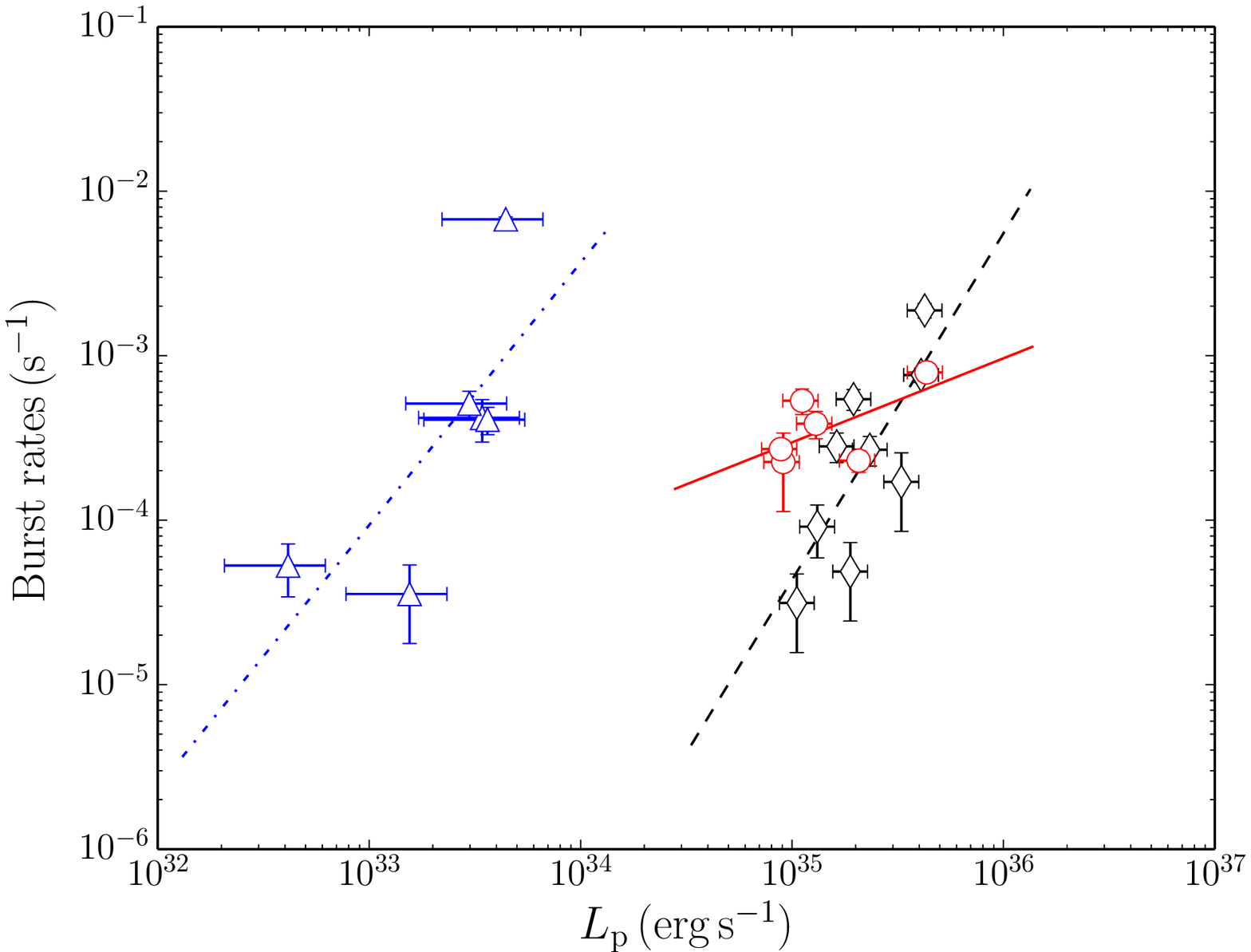}
\caption{Left panel, the relationship between the burst rate and the persistent luminosity. Right panel, three power laws are fitted for SGR 1806-20, 1E 1048-5937 and SGR 0501+4516, shown as dashed line, solid line and dot-dashed line, respectively.}
\label{freq}
\end{figure}

\reffig{freq} shows a scatter relation between the burst rate and the persistent flux for all the sources. The power law indices and correlation coefficients for \object{SGR 1806-20}, \object{1E 1048-5937} and \object{SGR 0501+4516} are $ 2.11 \pm 0.64,~ 0.51 \pm 0.33,~ 1.60 \pm 0.70 $ and $\rho = 0.78,~ 0.62,~ 0.75$, respectively. The relationship between the burst rate and $F_{\rm p}$ can be described as a power law with $\gamma = 1.19\pm 0.35$ and $\rho = 0.61$.

\begin{figure}
\centering
\includegraphics[width=0.4\textwidth]{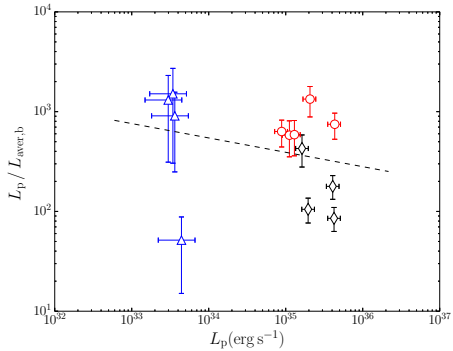}
\caption{In this figure, the symbols are the same as \reffig{stat}. It shows the correlation between the ratio ($L_{\rm p}/L_{\rm b}$) and the persistent luminosity. The dashed lines are the fitted power laws
described in Section 4.3.}
\label{ratio}
\end{figure}

We computed the ratio ($L_{\rm p}/L_{\rm aver,b}$), which could be rewritten as $E_{\rm p}/E_{\rm b}$, since

\begin{equation}
\frac{L_{\rm p}}{L_{\rm aver,b}} = \frac{L_{\rm p} \times t_{\rm p}}{L_{\rm b} \times t_{\rm b}} = \frac{E_{\rm p}}{E_{\rm b}}.
\end{equation}
The obtained ratios cover from several tens to several thousands, shown in \reffig{ratio}. We also fitted this relationship with a power law with index $\gamma = -0.14\pm 0.16$, and correlation coefficient is $-0.26$. These results indicate that the ratio ($L_{\rm p}/L_{\rm aver,b}$) has no or weak (negative) correlation with the radiation strength or sources. To estimate the statistical quantities, we also calculated the geometric mean and the geometric standard deviation. When we calculate these two value, a weighted statistics is considered and we used the reciprocal of error range as the weighted index for each observation. The geometric mean for this ratio is $361.51$, and the geometric standard deviation is $2.74$.

\section{Discussion}

In this paper, we showed the temporal and spectral analysis of short bursts in three AXPs/SGRs using the Bayesian blocks algorithm. The Bayesian blocks method checks each count recorded by the detector and determines whether it is a change point, which means that the time resolution for each block could reach the limit of the detector. Thus, the beginning and the end of each burst could also be determined in this precision, which makes it possible to analyze the duration of bursts precisely. We found the duration distributions for AXPs/SGRs can be fitted by the sum of two lognormal functions. Among all of these three sources, the mean values of the two components are at $\sim 0.1 \rm~ s$ and $\sim 1 \rm~ s$, respectively. Phenomenologically, one of this sources is dominated by longer ones, while the other two are dominated by shorter ones.  \citet{Govgucs2001} first showed the statistics of duration using \textit{RXTE} data and indicated the distributions peaked at $\sim 100~\rm ms$ for \object{SGR 1806-20} and \object{SGR 1900+14}. They also divided these short bursts into two components, named ``single pulse burst" and ``multi-peaked burst". These two components peak at $88.1~\rm ms$ and $229.9~\rm ms$ in \object{SGR 1806-20}, or peak at $46.7~\rm ms$ and $148.9~\rm ms$ in \object{SGR 1900+14}. The longer components are much shorter than our results ($\sim 1~\rm s$). This difference has two origins, the method to detect short bursts and the way to divide bursts into two classes. These results show that the Bayesian blocks algorithm has the ability to find bursts which are dim but long enough, and the existence of the long time scale tail of short bursts in AXPs/SGRs.

The ability of Bayesian blocks algorithm to find dim bursts is apparently affected by the count rates of the source. We regard the modeling of waiting time as a possible evidence to this conclusion. \citet{Cheng1996} first showed the waiting time distribution of SGR and compared it from with the cases in earthquake. \citet{Govgucs1999} and \citet{Govgucs2000} showed that the waiting time distribution may be fitted by a lognormal function for \object{SGR 1900+14} and \object{SGR 1806-20}. However, there is an unexpected bump in the short time scale of the distribution. They regarded this structure as a result of the uncertainty to determine the shape of bursts, which may make a multi-peaked burst become several single pulse bursts with shorter waiting times. In our results, \object{SGR 1806-20} and \object{1E 1048-5937} also showed the similar phenomenon, while \object{SGR 0501+4516} showed a better lognormal distribution. Comparing these two results, we regard the undetectable weak bursts as the reason for divergences in distribution, which is notable in \object{SGR 1806-20} and \object{1E 1048-5937} because of the low count rates. \object{SGR 0501+4516} is the nearest one among these three sources and it was also in its most luminous phase, which make it easy to detect the dim bursts. In this case, we attribute the differences of the three sources in our samples to the undetected weak bursts in \object{SGR 1806-20} and \object{1E 1048-5937}. Of course, the possibility can not be ruled out that our samples are completed in this energy band ($1-10 ~\rm keV$) for \object{SGR 1806-20} and \object{1E 1048-5937}, which do not have weaker bursts. However, the possibility is quite limited, considering the fact that the count rates for these two sources are only $\sim1~\rm cte s^{-1}$, and that waiting times we got in \object{SGR 1806-20} are $\sim 10$ times longer than the ones in \citet{Govgucs2000}.

The spectra were also analyzed for long and short time scale bursts, using the first observation from \object{SGR 0501+4516}. In our results, this observation contains the most burst photons and could be divided easily into two components with little interlock. We chose two models, two black bodies and OTTB, in our burst spectra fitting. Two black bodies model is one of the simplest model and widely used in burst spectra fitting \citep{Feroci2004}. \citet{Olive2004} analyzed an intermediate burst from \object{SGR 1900+14} and found that two black bodies model could provide an acceptable fit to both time resolved spectra and integrated spectrum. They attributed the higher temperature component to multi-temperature trapped fireball and regarded the lower one as the emission from star surface. \citet{Israel2008} suggested that the higher temperature component came from the surface of neutron star, while the lower one was emitted from a magnetospheric region. Ignoring the mechanism, double black bodies model could provide acceptable fits with reduced $\chi^2 \sim 1.1$ to the burst spectra for both shorter and longer burst in our work. We also used an alternative model (OTTB) to fit the burst spectra. This model is also widely used in the burst spectra fitting in AXPs/SGRs, but it is not always effective \citep{Feroci2004, Olive2004}. However, it works well in our burst spectra fitting too. Thus, we examined the chosen burst spectra using two universal models and both models works well judged by reduced $\chi^2 \sim 1.1$. These two models reflect different physical processes and we can not claim which is the truth on the surface of AXPs/SGRs. Fortunately, in both of these two models, the characteristic parameters, black body temperature or plasma temperature, show negligible variety in the error range. In that case, we prefer to regard that the two classes bursts we divided originate from the same resource, but how could the two time scale bursts be generated is still an issue to be considered.

The relationships between the short bursts and the persistent emission were analyzed to find hints for the energy origin of AXPs/SGRs. We show a power law with $\gamma = 1.23\pm0.18$ between the luminosity of persistent emission and burst. In accretion model, this phenomenon is natural, since the persistent radiation represents the accretion rate, while the burst radiation represents the consumption rate of the accreted matter. Considering the an equilibrium condition during an observation, the positive correlation is apparent between the accretion rate and the consumption rate, which results in the positive relationship between the luminosity of persistent emission and burst. In the magnetar model, this phenomenon is also natural. Both the persistent emission and the bursts are from the magnetic energy. During an outburst, some seismic activities may trigger magnetic reconnections or crystal fractures, which are responsible for the short bursts \citep{Thompson1996}. During this process, the magnetosphere will become more twisted. The corresponding persistent flux will also increase \citep{Beloborodov2009}.

We also introduced the energy ratio ($L_{\rm p}/L_{\rm aver,b}$) from Type \uppercase\expandafter{\romannumeral1} X-ray bursts to judge the energy release in AXPs/SGRs. In Type \uppercase\expandafter{\romannumeral1} X-ray bursts, the range of this ratio covers from several tens to $\sim 1000$ and does not vary with the persistent luminosity \citep{Galloway2008}. This character is regarded as a strong evidence for the nuclear burning model. We show that the energy ratios in AXPs/SGRs have the similar statistic character with the ones in Type \uppercase\expandafter{\romannumeral1} X-ray bursts. Considering that burst fluxes we got are lower limits, the energy ratio we got is the upper limit of the real one. The energy ratios in our sample cover from $\sim 10$ to $\sim 2000$, which is comparable with in Type \uppercase\expandafter{\romannumeral1} X-ray bursts. However, the nuclear burning model is not so suitable for AXPs/SGRs based on two reasons. On one hand, AXPs/SGRs are isolated stars without apparent accretion. On the other hand, the time scale of short bursts is much shorter than the prediction in nuclear burning model. In Type \uppercase\expandafter{\romannumeral1} X-ray bursts, the energy ratio can be calculated for each burst, while in AXPs/SGRs, this ratio can only be analyzed for a long period with many bursts. Although the ratios in Type \uppercase\expandafter{\romannumeral1} X-ray bursts and AXPs/SGRs are different, they both show that there should be a connection between the energy origin of persistent radiation and the resource of bursts. Nevertheless, we notice that there is no relevant prediction about this ratio in AXPs/SGRs models yet. We regard that this ratio ($L_{\rm p}/L_{\rm aver,b}$) may reflect some essence like in Type \uppercase\expandafter{\romannumeral1} X-ray bursts and should be involved into consideration in a successful model.

We would like to thank the pulsar group of PKU for helpful discussions and comments, and Hao Tong from XAO of CAS for supplements about magnetar model. We also thank Kathryn Plant and Laura Lopez for revising the whole manuscript. This research is based on data and software provided by the the ESA \textit{XMM-Newton} Science Archive (XSA) and the NASA/GSFC High Energy Astrophysics Science Archive Research Center (HEASARC). This work is supported by the 973 program (No. 2012CB821801), the National Natural Science Foundation of China (Grant Nos. 11225314, Grant Nos. 11133002, NSFC-11103020 and NSFC-11473027), National Found for Fostering Talents of Basic Science (No. J0630311), and XTP project XDA04060604, XDB09000000 (Supported by the Strategic Priority Research Program ``The Emergence of Cosmological Structures" of the Chinese Academy of Sciences, Grant No. XDB09000000). Z.S. Li is supported by China Postdoctoral Science Foundation (2014M560844).
\bibliographystyle{apj}

\newpage

\begin{table}
\begin{center}
\caption{The parameters of sources.}
\label{param}
\begin{tabular}{cccccccc}
\tableline
$\rm Source$ & $\rm Period$ & $ \rm Epoch$ & $\rm Ref.$ & $\rm Distance$ & $\rm Ref.$ & $N_{\rm H}$ & $\rm Ref.$ \\
& $\rm (s)$ & $\rm (MJD)$ & & $\rm (kpc) $ & &$(10^{22} \rm cm^{-2})$ & \\
\tableline
\object{SGR 1806-20} & $7.6022(7)$ & 54189 & $ (1) $ & ${8.7_{-1.5}^{+1.8}}$ & $(4)$ & $6.9\pm0.4$ & $(7)$\\
\object{1E 1048-5937} & $6.4578754(25)$ & 54185.9 & $(2)$ & $9.0\pm1.7$ & $(5)$ & $0.97\pm0.01$ & $ (8) $\\
\object{SGR 0501-4516} & $5.76209653(3)$ & 54750 & $(3)$ & $0.8\pm0.4$ & $(6)$ & ${0.6_{-0.3}^{+0.5}}$ &$(9)$ \\
\tableline
\end{tabular}
\tablerefs{(1)\citet{Nakagawa2009}, (2)\citet{Dib2009}, (3)\citet{Govgucs2010}, (4)\citet{Bibby2008}, (5)\citet{Durant2006}, (6)\citet{Leahy2007}, (7)\citet{Mori2013}, (8)\citet{Tam2008}, (9)\citet{Rea2009}}
\end{center}
\end{table}

\begin{table}
\begin{center}
\caption{The non-piled-up short bursts results. }
\label{shortBurst}
\begin{tabular}{cccc}
\tableline
Source & OBS-ID & $N_{\rm photon}$ & $N_{\rm burst}$\tablenotemark{a}\\
\tableline
\multirow{14}{*}{\object{SGR 1806-20}} & 0148210101 & 0 & 0(0) \\
& 0148210401 & 0 & 0(0) \\
& 0164561101 & 625 & 24(28) \\
& 0164561301 & 0 & 0(1)  \\
& 0164561401 & 47 & 6(6)  \\
& 0205350101 & 447 & 21(27) \\
& 0406600301 & 24 & 1(1)  \\
& 0406600401 &131 & 10(12)  \\
& 0502170301 & 47 & 5(6)  \\
& 0502170401 & 16 & 1(2)  \\
& 0554600301 & 0 & 0(0)  \\
& 0554600401 & 0 & 0(0)  \\
& 0604090201 & 0 & 0(0)  \\
& 0654230401 & 5 & 1(1)  \\
\hline
\multirow{5}{*}{\object{1E 1048-5937}} & 0112780401 & 14 & 1(1) \\
& 0147860101 & 155 & 8(11)\\
& 0307410201 & 77 & 7(7)\\
& 0307410301 & 87 & 8(8)\\
& 0510010601 & 467 & 24(25)\\
& 0654870101 & 264 & 21(21)\\
\hline
\multirow{10}{*}{\object{SGR 0501+4516}} & 0552971101 & 126 & 6(7)\\
& 0552971201 & 67 & 3(3)\\
& 0552971301 & 139 & 7(7)\\
& 0552971401 & 18 & 1(1)\\
& 0560191501 & 7214 & 218(270)\\
& 0560191501[1]\tablenotemark{b} & 1758 & 56(72)\\
& 0560191501[2]\tablenotemark{b} & 2983 & 70(82)\\
& 0560191501[3]\tablenotemark{b} & 1676 & 54(57)\\
& 0560191501[4]\tablenotemark{b} & 1123 & 44(56)\\
& 0604220101 & 20 & 2(2)\\
\tableline
\end{tabular}
\tablenotetext{a}{The raw burst (including piled-up ones) are listed in brackets.}
\tablenotetext{b}{These four segments labeled as 1 to 4 are divided from the observation ID 0560191501.}
\end{center}
\end{table}

\begin{table}
\begin{center}
\caption{ The spectral results of SGR 1806-20.}
\label{spec1806}
\begin{tabular}{ccccccc }
\tableline
\multirow{3}{*}{OBS-DATE} & \multirow{3}{*}{OBS-ID} & \multicolumn{3}{c}{\multirow{1}{*}{persistent spectrum}} & \multicolumn{2}{c}{\multirow{1}{*}{burst spectrum}}\\
& & \multicolumn{3}{c}{phabs(BB+PL)} & \multicolumn{2}{c}{phabs(BB)} \\
& & $kT (\rm keV)$ & $ \gamma $ & ${\rm red.}\chi^2 {\rm (d.o.f.)} $ & $kT (\rm keV) $ & ${\rm red.}\chi^2 {\rm (d.o.f.)} $ \\
\tableline

2003-04-03 & 0148210101 & $ {0.44^{+0.13}_{- 0.15 }}$ & $ {1.44^{+0.18}_{- 0.24 }}$ & $ 0.87 ( 100 )$ & & \\
2003-10-07 & 0148210401 & $ {0.57^{+0.08}_{- 0.11 }}$ & $ {1.37^{+0.20}_{- 0.24 }}$ & $ 1.10 ( 150 )$ & & \\
2004-10-06 & 0164561101 & $ {0.62^{+0.08}_{- 0.11 }}$ & $ {1.40^{+0.13}_{- 0.14 }}$ & $ 0.86 ( 263 )$ & $ {2.14^{+0.30}_{- 0.23 }}$ & $ 0.95 ( 56 )$ \\
2005-03-07 & 0164561301 & $ {0.70^{+0.07}_{- 0.09 }}$ & $ {1.22\pm 0.07}$ & $ 1.08 ( 135 )$ & & \\
2005-10-04 & 0164561401 & $ {0.65^{+0.04}_{- 0.05 }}$ & $ {1.38\pm 0.04}$ & $ 1.15 ( 259 )$ & & \\
2004-09-06 & 0205350101 & $ {0.66^{+0.05}_{- 0.06 }}$ & $ {1.27\pm 0.09}$ & $ 0.99 ( 325 )$ & $ {2.24^{+0.48}_{- 0.33 }}$ & $ 0.87 ( 39 )$ \\
2006-04-04 & 0406600301 & $ {0.63^{+0.04}_{- 0.05 }}$ & $ {1.23\pm 0.05}$ & $ 1.11 ( 228 )$ & & \\
2006-09-10 & 0406600401 & $ {0.63^{+0.06}_{- 0.07 }}$ & $ {1.51^{+0.15}_{- 0.17 }}$ & $ 1.05 ( 243 )$ & $ {2.80^{+2.57}_{- 0.85 }}$ & $ 0.26 ( 8 )$ \\
2007-09-26 & 0502170301 & $ {0.62^{+0.05}_{- 0.06 }}$ & $ {1.59^{+0.18}_{- 0.20 }}$ & $ 1.02 ( 211 )$ & & \\
2008-04-02 & 0502170401 & $ {0.60^{+0.05}_{- 0.06 }}$ & $ {1.55\pm 0.06}$ & $ 1.16 ( 196 )$ & & \\
2008-09-05 & 0554600301 & $ {0.54^{+0.04}_{- 0.05 }}$ & $ {1.58\pm 0.04}$ & $ 1.19 ( 229 )$ & & \\
2009-03-03 & 0554600401 & $ {0.53^{+0.05}_{- 0.07 }}$ & $ {1.60\pm 0.05}$ & $ 1.04 ( 203 )$ & & \\
2009-09-07 & 0604090201 & $ {0.52^{+0.04}_{- 0.05 }}$ & $ {1.57\pm 0.06}$ & $ 0.91 ( 176 )$ & & \\
2011-03-23 & 0654230401 & $ {0.52^{+0.04}_{- 0.05 }}$ & $ {1.52\pm 0.05}$ & $ 0.92 ( 201 )$ & & \\

\tableline
\end{tabular}
\end{center}
\end{table}

\begin{landscape}
\begin{table}[p]
\begin{center}
\caption{The spectral results of 1E 1048-5937.}
\label{spec1048}
\begin{tabular}{ccccccccc }
\tableline
\multirow{3}{*}{OBS-DATE} & \multirow{3}{*}{OBS-ID} & \multicolumn{4}{c}{\multirow{1}{*}{persistent spectrum}} & \multicolumn{2}{c}{\multirow{1}{*}{burst spectrum}}\\
& & \multicolumn{4}{c}{phabs(BB+BB+PL)} & \multicolumn{2}{c}{phabs(BB)} \\
& & $kT_{\rm 1} (\rm keV) $ &$kT_{\rm 2} (\rm keV) $ & $ \gamma $ & ${\rm red.}\chi^2 {\rm (d.o.f.)} $ & $kT (\rm keV) $ & ${\rm red.}\chi^2 {\rm (d.o.f.)} $ \\
\tableline
2000-12-28 & 0112780401 & $ {0.63\pm 0.05}$ & & $ {2.90^{+0.21}_{- 0.16 }}$ & $ 1.01 ( 79 )$ & \\
2003-06-16 & 0147860101 & $ {0.56\pm 0.03}$ & $ {0.95^{+0.11}_{- 0.14 }}$ & $ {3.58^{+0.34}_{- 0.24 }}$ & $ 1.13 ( 225 )$ & $ {0.65^{+0.21}_{- 0.14 }}$ & $ 0.61 ( 11 )$ \\
2005-06-16 & 0307410201 & $ {0.46^{+0.12}_{- 0.15 }}$ & $ {0.75^{+0.21}_{- 0.10 }}$ & $ {3.36^{+2.01}_{- 0.21 }}$ & $ 0.92 ( 165 )$ & $ {0.48^{+0.16}_{- 0.11 }}$ & $ 0.90 ( 10 )$ \\
2005-06-28 & 0307410301 & $ {0.37^{+0.06}_{- 0.05 }}$ & $ {0.72^{+0.05}_{- 0.04 }}$ & $ {6.09^{+2.04}_{- 1.40 }}$ & $ 1.22 ( 149 )$ & $ {0.62^{+0.29}_{- 0.18 }}$ & $ 1.56 ( 12 )$ \\
2007-06-14 & 0510010601 & $ {0.54\pm 0.04}$ & $ {0.88^{+0.05}_{- 0.06 }}$ & $ {3.44^{+0.38}_{- 0.23 }}$ & $ 1.38 ( 250 )$ & $ {0.59^{+0.06}_{- 0.05 }}$ & $ 0.91 ( 37 )$ \\
2011-08-06 & 0654870101 & $ {0.43^{+0.05}_{- 0.04 }}$ & $ {0.78^{+0.04}_{- 0.03 }}$ & $ {4.54^{+0.84}_{- 0.56 }}$ & $ 1.01 ( 208 )$ & $ {0.46^{+0.06}_{- 0.05 }}$ & $ 0.38 ( 20 )$ \\

\tableline
\end{tabular}
\end{center}
\end{table}
\end{landscape}

\begin{landscape}
\begin{table}[p]
\begin{center}
\caption{ The spectral results of SGR 0501+4516.}
\label{spec0501}
\begin{tabular}{ccccccccc }
\tableline
\multirow{4}{*}{OBS-DATE} & \multirow{4}{*}{OBS-ID} & \multirow{2}{*}{$N_{\rm H}$} &\multicolumn{3}{c}{\multirow{1}{*}{persistent spectrum}} & \multicolumn{3}{c}{\multirow{1}{*}{burst spectrum}}\\
& & &\multicolumn{3}{c}{phabs(BB+PL)} & \multicolumn{3}{c}{phabs(BB+BB)} \\
& & \multirow{2}{*}{$(10^{22}~\rm cm^{-2})$} & $kT$ & $ \gamma $ & ${\rm red.}\chi^2  $ & $kT_{\rm 1}$ & $kT_{\rm 2}$ & ${\rm red.}\chi^2 $ \\
& & & $(\rm keV)$ & & ${\rm (d.o.f.)}$ & $(\rm keV)$ & $(\rm keV)$ & $ {\rm (d.o.f.)}$\\
\tableline

2008-08-29 & 0552971101 & $ {0.85\pm 0.06}$ & $ {0.70\pm 0.01}$ & $ {2.91^{+0.09}_{- 0.10 }}$ & $ 1.15 ( 231 )$ & $ {0.58^{+0.24}_{- 0.16 }}$ & & $ 0.86 ( 8 )$ \\
2008-08-31 & 0552971201 & $ {0.86^{+0.09}_{- 0.10 }}$ & $ {0.71\pm 0.02}$ & $ {2.91^{+0.16}_{- 0.17 }}$ & $ 1.17 ( 187 )$ & $ {0.71^{+0.19}_{- 0.15 }}$ & & $ 1.59 ( 9 )$ \\
2008-09-02 & 0552971301 & $ {0.81^{+0.07}_{- 0.08 }}$ & $ {0.69\pm 0.01}$ & $ {2.94\pm 0.14}$ & $ 1.27 ( 211 )$ & $ {0.59^{+0.18}_{- 0.13 }}$ & & $ 1.09 ( 9 )$ \\
2008-09-30 & 0552971401 & $ {0.84\pm 0.06}$ & $ {0.66\pm 0.01}$ & $ {3.16\pm 0.11}$ & $ 1.03 ( 204 )$ & \\
2008-08-23 & 0560191501 & $ {0.87\pm 0.03}$ & $ {0.70\pm 0.01}$ & $ {2.76\pm 0.05}$ & $ 1.06 ( 292 )$ & $ {0.57\pm 0.04}$ & $ {2.13^{+0.28}_{- 0.21 }}$ & $ 1.10 ( 212 )$ \\
2008-08-23 & 0560191501[1]\tablenotemark{a} & $ {0.84^{+0.06}_{- 0.07 }}$ & $ {0.70\pm 0.02}$ & $ {2.72^{+0.10}_{- 0.11 }}$ & $ 1.03 ( 224 )$ & $ {0.50^{+0.13}_{- 0.12 }}$ & $ {1.51^{+0.49}_{- 0.22 }}$ & $ 1.03 ( 58 )$ \\
 & 0560191501[2]\tablenotemark{a} & $ {0.81\pm 0.07}$ & $ {0.67\pm 0.02}$ & $ {2.62^{+0.10}_{- 0.11 }}$ & $ 1.01 ( 228 )$ & $ {0.67\pm 0.06}$ & $ {2.97^{+1.57}_{- 0.67 }}$ & $ 1.04 ( 97 )$ \\
 & 0560191501[3]\tablenotemark{a} & $ {0.90\pm 0.06}$ & $ {0.71\pm 0.02}$ & $ {2.83^{+0.09}_{- 0.10 }}$ & $ 0.99 ( 225 )$ & $ {0.52^{+0.10}_{- 0.09 }}$ & $ {1.66^{+0.71}_{- 0.30 }}$ & $ 1.00 ( 54 )$ \\
 & 0560191501[4]\tablenotemark{a} & $ {0.94\pm 0.06}$ & $ {0.71\pm 0.02}$ & $ {2.87\pm 0.09}$ & $ 1.22 ( 227 )$ & $ {0.55^{+0.14}_{- 0.22 }}$ & $ {2.16^{+-2.16}_{- 0.90 }}$ & $ 1.48 ( 35 )$ \\
2009-08-30 & 0604220101 & $ {0.93^{+0.12}_{- 0.11 }}$ & $ {0.54\pm 0.02}$ & $ {4.37^{+0.36}_{- 0.30 }}$ & $ 1.19 ( 122 )$ & \\

\tableline
\end{tabular}
\tablenotetext{a}{The same segmentation of observation ID 0560191501 described in \reftable{shortBurst}.}
\end{center}
\end{table}
\end{landscape}

\end{document}